\documentclass[aps,pra,epsf,twocolumn,showpacs]{revtex4}
\input{epsf}
\usepackage{epsfig}
\usepackage{graphicx}
\begin{document}

\title{Time-separated entangled light pulses from a single-atom emitter}

\author{David Vitali,$^{1}$ Priscilla Ca$\mathrm{\tilde{n}}$izares,$^{1}$
J\"urgen Eschner,$^{2}$ and Giovanna Morigi $^{3}$}

\affiliation{$^1$ Dipartimento di Fisica,
Universit\`a di Camerino, 62032 Camerino, Italy,\\
$^{2}$ ICFO - Institut de Ciencies Fotoniques,
Mediterranean Technology Park, 08860 Castelldefels (Barcelona), Spain\\
$^{3}$ Departament de Fisica, Universitat Autonoma de Barcelona,
08193 Bellaterra, Spain}

\date{\today}

\begin{abstract}
The controlled interaction between a single, trapped, laser-driven atom and the mode of a
high-finesse optical cavity allows for the generation of temporally separated, entangled
light pulses. Entanglement between the photon-number fluctuations of the pulses is
created and mediated via the atomic center-of-mass motion, which is interfaced with light
through the mechanical effect of atom-photon interaction. By means of a quantum noise
analysis we determine the correlation matrix which characterizes the entanglement, as a
function of the system parameters. The scheme is feasible in experimentally accessible
parameter regimes. It may be easily extended to the generation of entangled pulses at
different frequencies, even at vastly different wavelengths.
\end{abstract}


\maketitle

\section{Introduction}
\label{sec:intro}

The increasing interest in realizing quantum networks by means of
atoms and photons has risen the issue of achieving full coherent
control on atom-photon interactions. Photonic interfaces with atomic
ensemble, namely, with a macroscopic number of atoms, have been
explored in several milestone experiments, which demonstrated the
generation of single photon sources \cite{Kimble03}, two-mode
squeezing in the polarization of the emitted light \cite{Giacobino},
atomic memory for quantum states of light~\cite{LukinScience03,
Lukin03, Polzik04}, entanglement of remote
ensembles~\cite{Chou2005}, and teleportation between light and
matter~\cite{PolzikCirac06}.

Complementary to this approach, photonic interfaces using single
atoms can take advantage of the large level of control that can be
achieved on the atomic internal and external degrees of freedom.
For instance, in microwave cavity QED quantum state and
entanglement engineering using single atoms have been demonstrated
\cite{MicroCQED,CavityQED-Walther}. In the optical regime,
milestone experiments demonstrated the realization of one-atom
laser \cite{An94, Kimble-atomlaser}, revealed the mechanical
forces of single photons on single atoms \cite{Kimble-fly,
Rempe-fly, Bushev2005}, achieved the controlled interaction of a
trapped ion and a cavity \cite{Guthoehrlein01, Mundt02}, yielding
single-photon generation on demand \cite{Kuhn02, Kimble-photon,
Keller04}, and characterized the entanglement between a single
atom and its emitted photon~\cite{Monroe04, Weinfurter2005}. This latter step was instrumental for establishing entanglement between two distant trapped particles by projective measurement of the emitted photon~\cite{Moehring07}. Most recently, reversible quantum state transfer between light and atoms in a cavity have been experimentally demonstrated~\cite{Boozer07}. These
results constitute relevant progress towards the realization of
quantum networks with single atoms~\cite{CiracKimble, Kraus04}.

\begin{center} \begin{figure}[htb]
\includegraphics[width=0.55\textwidth]{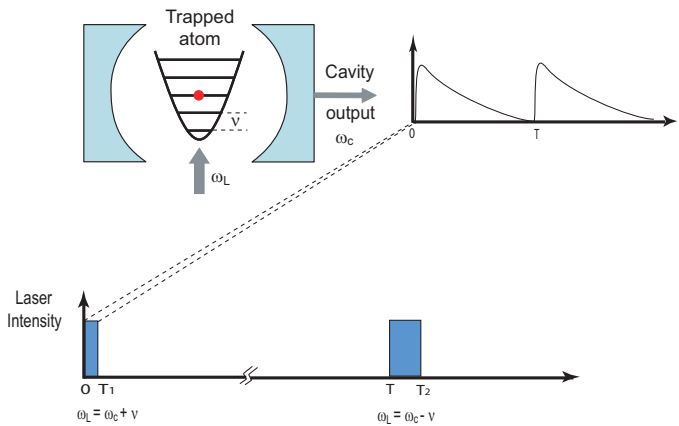}
\caption{A trapped atom is confined by a harmonic potential of frequency $\nu$ inside a
resonator. The atomic dipole is driven by two temporally separated laser pulses, whose
intensity as a function of time is displayed at the bottom, and couple with a cavity mode
at frequency $\omega_c$. The intensities of the two emitted pulses at the cavity output
are shown as a function of time. In this article we show that, in a suitable parameter
regime, they exhibit quadrature entanglement.} \label{Fig:0} \end{figure} \end{center}

Possible implementation of quantum networks with continuous variables~\cite{Bra04} using
single atoms as interfaces requires the controlled interaction of light with the atom
external degrees of freedom, which exploits the mechanical effects of light-atom
interactions~\cite{Ze-Parkins99,Parkins99,Parkins02}. Using these concepts, in a recent
proposal we predicted that a single atom, confined inside a resonator, can act as a
quantum medium, generating quantum correlations in the emitted light~\cite{PRL,Morigi06}.
Such correlations can be two-mode squeezing type of correlations, which for bipartite
Gaussian systems are synonymous of EPR-entanglement~\cite{Bra04,Reid}. In particular,
under suitable conditions, two classical light (laser) pulses, temporally separated at
the input, exhibit two-mode squeezing type of correlations at the output of this kind of
device, as sketched in Fig.~\ref{Fig:0}. In this case the quantum state of the atomic
motion serves as intermediate memory which mediates the entanglement between the first
and second pulse at the cavity output. Variation of the laser parameters, driving the
atom, allows for tuning the degree of entanglement between the pulses.

In this work, we analyze the efficiency of the proposal for
temporally-separated entangled pulses with single
atoms~\cite{PRL,Morigi06} by using a quantum Langevin equation
description. This description permits us to determine the
correlation matrix and hence the amount of entanglement one obtains
using experimentally accessible parameter regimes. We show that this
proposal is viable to existing experiments, hence providing an
important step towards continuous-variable photonic interfaces with
single atoms. We remark that here the atom acts as a source of
continuous variable ``time-bin'' entangled pulses, which could be an
alternative solution for secure communication \cite{Grang03} with
respect to those employing single-photon qubits~\cite{Gisetal02}.

This article is organized as follows. In Sec.~\ref{Sec:Theory} the
model determining the system's dynamics is introduced, and the
quantum Langevin Equations for the coupled dynamics between atomic
motion and cavity mode are derived. In
Sec.~\ref{Sec:Correlation:Matrix} the corresponding correlation
matrix for the two propagating correlated pulses is determined and
the degree of entanglement characterized. In Sec~\ref{Sec:Results}
entanglement is discussed as a function of the experimental
parameters. The conclusions and outlooks are presented in
Sec.~\ref{Sec:Conclusions}, and the appendices report the details
of the derivations in Sec.~\ref{Sec:Theory} and
Sec.~\ref{Sec:Correlation:Matrix}.

\section{Theoretical description} \label{Sec:Theory}

\begin{center} \begin{figure}[htb]
\includegraphics[width=0.5\textwidth]{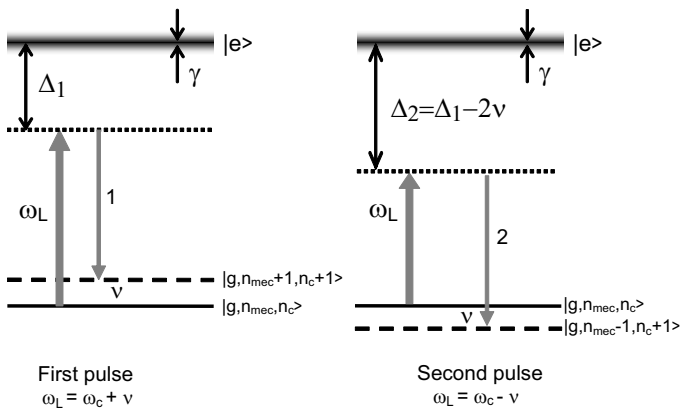}
\caption{Relevant energy level and resonant transitions during the two pulses. Here,
$|g\rangle$ and $|e\rangle$ denote ground and excited state of the atomic dipole
transition with linewidth $\gamma$, $|n_c\rangle$ the number of cavity photons, and
$|n_{\rm mec}\rangle$ the number of vibrational excitations. (a) First pulse: The laser
is detuned by $\Delta_1$ from the dipole, and by $\delta_1=\nu$ from the cavity mode,
driving resonantly the two-photon transition $|g,n_{\rm mec},n_c\rangle\to|g,n_{\rm
mec}+1,n_c+1\rangle$. These processes generate two-mode squeezing between cavity and
motion. (b) Second pulse: The laser is detuned by $\Delta_2=\Delta_1-2\nu$ from the
dipole, hence driving resonantly the two-photon transition $|g,n_{\rm
mec},n_c\rangle\to|g,n_{\rm mec}-1,n_c+1\rangle$. Choosing the pulse duration properly,
quantum-state transfer between motion and cavity mode can be achieved,
see~\cite{Ze-Parkins99,Parkins99}. Note that by varying the frequency of the second laser
pulse and/or the atomic transition which it excites, the two output pulses can be at
different frequencies or wavelengths.} \label{Fig:1}
\end{figure}
\end{center}

We summarize the basic concepts at the basis of the proposal for the creation of pairs of
temporally separated, continuous-variable (CV) entangled pulses. Let us consider a single
atom in a harmonic trap of frequency $\nu$ and confined inside a resonator, in the setup
sketched in Fig.~\ref{Fig:0}. Be $b$, $b^\dagger$ the annihilation and creation operator
of an excitation of the quantized motion inside the trap. The atomic dipole couples with
a cavity mode at frequency $\omega_c$, which is far-off resonance from the dipole
frequency $\omega_0$. We denote by $a$, $a^\dagger$ the annihilation and creation
operator of a cavity photon. A first laser pulse at frequency $\omega_L \simeq
\omega_c+\nu$ illuminates the atom in the time interval $[0,T_1]$. In this regime, the
simultaneous emission of a cavity photon and of a vibrational quantum of the atom is
resonantly enhanced, see Fig.~\ref{Fig:1}(a), and the relevant dynamics are described the
effective Hamiltonian \begin{equation} H^{(1)}={\rm i}\hbar\chi_1
a^{\dagger}b^{\dagger}+{\rm
 H.c.}, \label{H:1:1}
\end{equation} where $\chi_1$ is the transition amplitude of the
resonant process. This Hamiltonian describes an interaction giving rise to two-mode
squeezing, i.e., CV entanglement between the center-of-mass oscillator and the cavity
mode. If the pulse is implemented for a sufficiently short time, so that photon leakage
out of the cavity has negligible effects and the dynamics can be assumed to be coherent,
at the end of the pulse the cavity mode and the atom's vibrational motion will be
entangled. Cavity decay will give rise to a propagating pulse, whose photon number is
quantum-correlated with the vibrational phonon number~\cite{Parkins99,Parkins02}.

We then assume that at a time $T >T_1$ a second laser pulse tuned
to the frequency $\omega_L \simeq \omega_c-\nu$ drives the atom
till the time $T_2$. In this regime, one has resonant enhancement of the emission of a cavity
photon with the simultaneous \emph{absorption} of a vibrational quantum, see
Fig.~\ref{Fig:1}(b). The effective dynamics is described by
the Hamiltonian
\begin{equation} H^{(2)}={\rm i}\hbar\chi_2
a^{\dagger}b+{\rm H.c}.,\label{H:2:1}
\end{equation}
where $\chi_2$ is the transition amplitude of the resonant process. Again, we assume that
cavity decay can be safely neglected, so that the dynamics is coherent. When the pulse
duration $\delta T=T_2-T$ is appropriately chosen, the quantum state of the
center-of-mass motion at time $T$ is completely transferred to the cavity mode at time
$T_2$. Consequently, by cavity decay a second pulse at the cavity output will be
generated, which is entangled with the first one~\cite{PRL,Morigi06}.

This proposal is based on the assumption that during the laser pulses the dynamics is
essentially described by the effective Hamiltonian~(\ref{H:1:1}) and~(\ref{H:2:1}), while
detrimental effects like atomic spontaneous emission, vacuum optical input noise entering
the cavity, and fluctuations of the trapping potential can be neglected. These
assumptions are justified in certain parameter regimes, which have been discussed
in~\cite{PRL,Morigi06}. The scope of this work is to explore the robustness of the scheme
when detrimental effects are small but cannot be {\it a priori} neglected in the
dynamical equations. At this purpose, in this section we adopt a quantum Langevin
equations treatment, taking into account all sources of noise. This permits us to
determine the correlation matrix for the generated pulses, which are discussed in
Sec.~\ref{Sec:Correlation:Matrix}, and to quantify their degree of entanglement for a
wide range experimental parameters, as shown in Sec~\ref{Sec:Results}.

\subsection{The system}

We consider an atom of mass $m$, whose center-of-mass motion takes
place essentially in one-dimension. We assume, in fact, that the
radial potential is sufficiently steep, so that the radial motion
can be considered frozen out. Be the motion along the
$\hat{x}$-axis, and be $x,p$ the position and momentum of the
atomic center of mass. The center of mass is a harmonic oscillator
with angular frequency $\nu$, whose Hamiltonian reads
\begin{equation} \label{H:mec} H_{\rm
mec}=\hbar\nu\left(b^{\dagger}b+\frac{1}{2}\right)~,
\end{equation} where $x,p$ are related to the annihilation and
creation operators $b$ and $b^{\dagger}$ of a quantum of
vibrational energy $\hbar\nu$ by the relations
$x=\sqrt{\hbar/2m\nu}(b+b^{\dagger})$ and $p={\rm i}\sqrt{\hbar
m\nu/2}(b^{\dagger}-b)$. The relevant atomic internal degrees of
freedom are the ground state $|g\rangle$ and the excited state
$|e\rangle$, which form a dipole transition with moment ${\bf d}$
and frequency $\omega_0$. The transition couples to one optical
mode of the cavity at frequency $\omega_c$ and to the laser, a
classical field, which is a pulsed excitation whose central
frequency $\omega_L$ is tuned from a pulse to the next. We define
our model in the reference frame rotating at $\omega_L$,
remembering that the frequency changes from the first to the
second pulse, and the two reference frames are hence related by a
global time-dependent phase. In the reference frame rotating at
$\omega_L$ the total Hamiltonian is $H=H_a+H_c+H_{\rm int}$. Here,
\begin{equation} H_a=-\hbar\Delta|e\rangle\langle e|+H_{\rm mec}
\label{H:atom} \end{equation} is the atomic Hamiltonian with
$\Delta=\omega_L-\omega_0$, while \begin{equation}
H_c=-\hbar\delta a^{\dagger}a \end{equation} describes the
dynamics of the cavity mode, with $\delta=\omega_L-\omega_c$. The
coupling between atom and fields is described by \begin{equation}
H_{\rm int}=\hbar\left(\sigma^{\dagger}B(t)+\sigma
B^{\dagger}(t)\right)~, \end{equation} where $\sigma =
|g\rangle\langle e|$ and $\sigma^{\dagger} = |e\rangle\langle g|$
denote the dipole lowering and raising operators and $B(t)$ is the
operator for the field degrees of freedom, which we decompose into
laser and cavity components, $B(t)=B_L(t)+B_c$. The cavity term is
$B_c=g_c \cos(k x\cos\theta_c+\phi_c) a$, where $g_c$ is the
coupling strength and the cavity mode wave vector $\vec{k}$
($k=|\vec{k}|$) forms an angle $\theta_c$ with the axis $\hat{x}$
of the motion. The angle $\phi_c$ takes into account the position
of the trap center inside the cavity. From now on we will assume
that the atomic motion is in the Lamb-Dicke regime, such that the
atom-photon interactions can be expanded at second order in the
Lamb-Dicke parameter $\eta=k\sqrt{ \hbar/2m\nu}$. In this limit
term $B_c$ takes the form~\cite{Footnote} \begin{eqnarray} B_c
&=&g_c \cos\phi_c a\left(1-\frac{\eta^2}{2}\cos^2 \theta_c (2
  b^{\dagger}b+1)\right)\nonumber\\
  &&-\eta \cos\theta_c g_c \sin\phi_c a (b^{\dagger}+b).
  \label{b_c}
\end{eqnarray}
The laser term is $B_L(t)=\Omega(t){\rm e}^{{\rm
i}kx\cos\theta_L}$, where $\Omega (t)$ is the (slowly-varying) Rabi
frequency and $\theta_L$ is the angle between the direction of
propagation of the laser and the trap axis. In the Lamb-Dicke regime
this operator reads
\begin{eqnarray}B_L(t)&=&\Omega(t)\left(1-\frac{\eta^2}{2}\cos^2
\theta_L (2 b^{\dagger}b+1)\right)\\
&+&i\eta\Omega(t)\cos\theta_L (b^{\dagger}+b). \nonumber
\label{b_L} \end{eqnarray}

\subsection{Quantum Langevin equations} \label{Sec:QLE}

The full dynamics of the system must take into account the coupling with the environment,
which is here represented by the dipole fluctuations, giving rise to spontaneous
emission, by vacuum fluctuations at the cavity input, giving rise to cavity decay, and by
fluctuations of the trapping potentials, which are responsible of damping and loss of
quantum coherence of the center-of-mass motion. Cavity decay is described by the
Markovian noise operator $a^{in}(t)$, while the heating due to the fluctuations of the
trap potential is described by the phenomenological Markovian input noise operator
$b^{in}(t)$ acting on the atomic motion. These two noise sources are mutually
uncorrelated and have zero mean value, and their only nonzero second-order correlation
functions are
\begin{eqnarray}
&&\langle a^{in}(t)a^{in}(t')^{\dagger}\rangle = \delta(t-t'), \\
&& \langle b^{in}(t)b^{in}(t')^{\dagger}\rangle = \left(\bar{N}+1\right)\delta(t-t'), \\
&& \langle b^{in}(t)^{\dagger} b^{in}(t')\rangle =\bar{N} \delta(t-t'),
\end{eqnarray}
where $\bar{N}$ is mean thermal vibrational number of the effective thermal reservoir
coupling to the atom center-of-mass motion. We denote by $\kappa_b$ the damping rate of
the vibrational motion. Pure heating corresponds to the limit of $\kappa_b \to 0$,
simultaneously with an infinite temperature of the associated effective reservoir, i.e.,
$\bar{N} \to \infty$, with $ 2\kappa_b\bar{N} \equiv \kappa_h$, the heating rate, kept
constant~\cite{Milburn-Schneider}.

The input noise terms associated with spontaneous emission are more involved because the
latter affects both the internal and the motional degree of freedom of the atom, due to
the presence of recoil. This effect has been neglected in the analysis presented
in~\cite{Vitali06}, which focussed on the c.w.-generation of entangled light, and will be
systematically taken into account in this work. At this purpose, the corresponding
Langevin force must be defined. At second order in the Lamb-Dicke parameter $\eta$, the
Langevin force operator $F(t)$ associated with spontaneous emission is given by (see also
its detailed derivation in App. A)
\begin{eqnarray}\label{eq:bigF}
F(t)&=&\sqrt{\gamma}\int {\rm d}\cos\theta \sqrt{{\cal N}(\cos\theta)} f_{\theta}^{in}(t)\\
& &\times \left(1-\frac{\eta^2}{2}\cos^2\theta(2b^{\dagger}b+1)+i\eta\cos\theta(b^{\dagger}+b)\right),\nonumber
\end{eqnarray}
where $\gamma$ is the spontaneous emission rate of level $|e\rangle$, and ${\cal
N}(\cos\theta)$ is the dipole pattern of emission, such that $\int {\rm d}\cos\theta
{\cal N}(\cos\theta)=1$. The integral is over the angle $\theta$ between the wave vector
of the emitted photon and the axis of the motion, and $f_{\theta}^{in}(t)$ are the
zero-mean, angle-dependent Langevin forces, with the only nonzero correlation function
\begin{equation}
\langle f^{in}_{\theta}(t)f^{in\dagger}_{\theta'}(t')\rangle=\delta(\theta-\theta')\delta(t-t').
\end{equation}
It is convenient to isolate the zeroth-order term in the Lamb-Dicke parameter in
Eq.~(\ref{eq:bigF}) and to rewrite $F(t)$ as
\begin{equation}\label{eq:bigF2}
F(t)=f^{in}(t)+ F_{nl}(t),
\end{equation}
with the zeroth-order term \begin{equation} \label{eq:finsem}
f^{in}(t)=\int {\rm d}\cos\theta \sqrt{{\cal N}(\cos\theta)} f^{in}_{\theta}(t)
\end{equation}
possessing the correlation function
\begin{equation} \langle f^{in}(t)f^{in\dagger}(t')\rangle=\delta(t-t').
\end{equation}
Term $F_{nl}(t)=F(t)-f^{in}(t)$ is hence at higher order in $\eta$. We shall see that
spontaneous emission noise is essentially described by $f^{in}(t)$, because the higher
order term $F_{nl}(t)$ gives a negligible contribution to the Langevin equations in the
parameter regime considered in this work (see App.~B).

In the frame rotating at the laser frequency the quantum Heisenberg--Langevin equations
(HLE) of the system read~\cite{milwal}
\begin{eqnarray}
\dot{a}(t)&=&(i\delta-\kappa)
a(t)+i\sigma(t)\left[B^{\dagger}(t), a(t)\right]
+\sqrt{2\kappa}a^{in}(t), \label{priqle}\\
\dot{b}(t)&=&-i\nu
b(t)+i \sigma(t)\left[B^{\dagger}(t)+F^{\dagger}(t), b(t)\right] \\
&& +i
\sigma(t)^{\dagger}\left[B(t)+F(t), b(t)\right]-\kappa_b b(t)+\sqrt{2\kappa_b}b^{in}(t), \nonumber \\
\dot{\sigma}(t)&=&\left[i\Delta -\frac{\gamma}{2}\right] \sigma(t)+i \sigma_z(t) \left( B(t)+F(t)\right), \label{penlastqle}\\
\dot{\sigma}_z(t)&=& 2i\sigma(t)\left[
B^{\dagger}(t)+F^{\dagger}(t)\right]-2i\sigma^{\dagger}(t)\left[
B(t)+F(t)\right]\nonumber \\
&&-\gamma \left(\sigma_z(t)+1\right)/2, \label{lastqle}
\end{eqnarray}
where $\sigma_z=\sigma^{\dagger}\sigma-\sigma
\sigma^{\dagger}$ and $\kappa$ is the decay rate of the cavity mode.

In the cases we are going to consider both laser as well as cavity mode are tuned far-off
resonance from the dipole transition, i.e., $|\Delta|\gg\Omega,g,|\delta|,\gamma$. In
this regime the atomic internal degrees of freedom can be eliminated in second order
perturbation theory. Hence, we neglect the time evolution of $\sigma_z$,
Eq.~(\ref{lastqle}), and approximate $\sigma_z(t)\approx -1$. Correspondingly,
Eq.~(\ref{penlastqle}) becomes
\begin{equation} \dot{\sigma}(t)=-\left(\frac{\gamma}{2}-i\Delta
\right)\sigma(t)-i B(t)- i F(t), \label{lastqle2} \end{equation} whose
formal solution is \begin{eqnarray}
&& \sigma(t)=e^{-\left(\frac{\gamma}{2}-i\Delta \right)t}\sigma(0) \label{lastqlesol} \\
&&-i\int_0^t {\rm d}s e^{-\left(\frac{\gamma}{2}-i\Delta \right)s}
\left[ B(t-s)+ F(t-s)\right]. \nonumber \end{eqnarray} We insert
solution~(\ref{lastqlesol}) into the other HLE and neglect the
transient term as we are interested in the dynamics at times which
are much larger than $1/|\Delta|$. The resulting HLE for cavity
and trap oscillators are \begin{widetext} \begin{eqnarray}
\label{HLE:0:1} \dot{a}^{\dagger}(t)&=&-i\delta
a^{\dagger}(t)-\int_0^t {\rm d}s~ e^{-\left(\frac{\gamma}{2}+i\Delta\right)s}
\left[B^{\dagger}(t-s)+F^{\dagger}(t-s)\right]\left[B(t), a^{\dagger}(t)\right] -\kappa a^{\dagger}(t)+\sqrt{2\kappa}a^{in\dagger}(t), \\
\dot{b}(t)&=&-i\nu b(t)+\int_0^t {\rm d}s~
e^{-\left(\frac{\gamma}{2}-i\Delta \right)s}
\left[B(t-s)+F(t-s)\right]\left[B^{\dagger}(t)+F^{\dagger}(t), b(t)\right] \nonumber \\
&&-\int_0^t {\rm d}s~ e^{-\left(\frac{\gamma}{2}+{\rm i}\Delta
\right)s} \left[B^{\dagger}(t-s)+ F^{\dagger}(t-s)\right]\left[B(t)+F(t), b(t)\right] -\kappa_b
b(t)+\sqrt{2\kappa_b}b^{in}(t), \label{HLE:0:2}
\end{eqnarray}
\end{widetext}
where we have not taken care of operator ordering, since, as we shall see, within the
validity limit of our treatment these integral terms will generate only linear
contributions. We now determine the solutions of these equations for the dynamics during
the first pulse, between the pulses, and during the second pulse.

\subsubsection{Dynamics during the first laser pulse}

We first consider the dynamics during the first pulse, i.e., in the time interval $0 \leq
t \leq T_1$, where we assume a square laser pulse with central angular frequency
$\omega_{L1}$ and constant Rabi frequency $\Omega(t)=\Omega_1$ during this time interval.
Correspondingly, we denote by $\Delta=\Delta_1$ and $\delta=\delta_1$ the detuning of the
laser from the atomic and cavity frequency. We assume that the laser is far-off resonance
from the atomic transition, i.e., $\Delta_1$ is negative and
$|\Delta_1|\gg\Omega_1,g,\gamma$. The laser frequency is tuned to the value $\omega_{L1}
\simeq \omega_c+\nu$, namely, the cavity mode is resonant with the Stokes motional
sideband of the laser light. This condition allows to establish a parametric-amplifier
type of interaction between cavity mode and motion, which is selectively enhanced
provided that the Stokes sideband is spectrally resolved, namely, when $\nu T_1\gg 1$ and
$\nu\gg\kappa,\kappa_b$. In particular, we take into account the presence of a.c.-Stark
shifts $\delta\nu$, due to the mechanical coupling with laser and cavity modes, by tuning
the laser frequency to the value
\begin{equation}
\label{eq:res1} \omega_{L1}=\omega_c+\nu^{\prime},
\end{equation}
and hence $\delta_1^{\prime}= \nu^{\prime}$, with $\nu^{\prime}=\nu+\delta\nu$ and
$\delta_1^{\prime}=\delta_1-\delta^{\prime}$, and $\delta^{\prime}$ accounts for possible
a.c.-Stark shifts due to off-resonant couplings. The value of $\delta\nu$ is determined
in a self-consistent way, which is extensively discussed in App.~\ref{App:A}, see
also~\cite{Vitali06}.

Starting from Eqs.~(\ref{HLE:0:1})-(\ref{HLE:0:2}), in this
parameter regime we derive the effective HLE, which describe the
coherent interaction between the cavity mode and the vibrational
motion during the first laser pulse, in the presence of losses and
noise processes due to spontaneous emission, cavity decay, and
vibrational heating. For later convenience, we study the equations
in the reference frame rotating at the cavity-mode frequency,
which is obtained from the reference frame of the laser frequency
by the transformation \begin{eqnarray} && \tilde{a}^{\dagger}(t)=
e^{i\nu 't} a^{\dagger}(t),\\&&\tilde{b}(t)= e^{i\nu 't} b(t).
\end{eqnarray}
In this reference frame, the HLE
read \begin{widetext} \begin{eqnarray} &&
\dot{\tilde{a}}^{\dagger}(t)=\chi_1^*\tilde{b}(t) -\left(\kappa
+\kappa_{L}\right)
\tilde{a}^{\dagger}(t)+\sqrt{2\kappa}\tilde{a}^{in\dagger}(t)
+\sqrt{2}\bar{\kappa}_{L}\tilde{a}_{-}^{in\dagger}(t)~, \label{qlefin1}\\
&& \dot{\tilde{b}}(t)=\chi_1 \tilde{a}^{\dagger}(t)
-\left(\kappa_b+\kappa_{+1}^{b}-\kappa_{-1}^{b}\right)
\tilde{b}(t)+\sqrt{2\kappa_b}\tilde{b}^{in}(t)
+\sqrt{2}\bar{\kappa}_{+1}^{b}\tilde{a}_{+}^{in}(t)-\sqrt{2}\bar{\kappa}_{-1}^{b}\tilde{a}_{-}^{in\dagger}(t)~.
\label{qlefin3} \end{eqnarray} \end{widetext} The equations are at
second order in the Lamb-Dicke parameter, and have been obtained
neglecting off-resonant terms. The steps of the derivation are
reported in App.~\ref{App:A}.

We now define and discuss each term appearing in the equations. The
effective coupling between motion and cavity mode is
\begin{eqnarray} &&\chi_1=\eta \frac{\Omega_1
g_c^*\cos\phi_c}{\Delta_1}\left(\cos\theta_L+{\rm i}\tan\phi_c\cos\theta_c\right)~,\label{Chi:1:app} \end{eqnarray} and corresponds to the Raman
processes in which laser photons are scattered into the cavity mode with a change in the center-of-mass excitation. The new noise operators
appearing in Eqs.~(\ref{qlefin1})-(\ref{qlefin3}), $\tilde{a}_{\pm}^{in}(t)$, are defined as
\begin{eqnarray}
&&\tilde{a}_{-}^{in}(t)=f^{in}(t)e^{-i\nu 't}, \label{nos1} \\
&&\tilde{a}_{+}^{in}(t)=f^{in}(t)e^{i\nu 't}, \label{nos2}
\end{eqnarray} and describe the coupling with external optical
modes due to the photons scattered by the atom. They possess the same correlation
functions of the spontaneous emission noise $f^{in}(t)$, and at the time scales of
interest, $\nu't \gg 1$, they are uncorrelated from each other. These noise components
affect both the cavity mode and the vibrational motion. In particular,
$\tilde{a}_{-}^{in}(t)$ describes quantum noise associated with incoherent scattering by
the atom of a cavity photon into the external modes at rate
\begin{equation}
\kappa_{L}=\frac{\gamma}{2}\frac{|g_c|^2\cos^2\phi_c}{\gamma^2/4+(\Delta_1-\nu')^2},
\label{los1}
\end{equation}
while the corresponding input noise scales with
\begin{equation}\label{los1bar}
\bar{\kappa}_{L}=-\sqrt{\frac{\gamma}{2}}\frac{g_c\cos\phi_c}{\gamma/2+{\rm
i}(\Delta_1-\nu')},
\end{equation}
where $\kappa_{L}=|\bar{\kappa}_{L}|^2$. The two noise terms in
Eqs.~(\ref{nos1})-(\ref{nos2}) also affect the atom's motion due to the mechanical
effects of the scattering of laser photons. The incoherent change of vibrational quanta
due to scattering of laser photons takes place at rates~\cite{Stenholm86}
\begin{equation}
\kappa_{\pm1}^{b}=
\eta^2\frac{\gamma}{2}\frac{|\Omega_1|^2\cos^2\theta_L}{\gamma^2/4+(\Delta_1\pm\nu')^2}
\label{Photon:scatter}~,
\end{equation}
while the corresponding terms $\bar{\kappa}_{\pm 1}^{b}$ scaling the input noise are
given by (see App.~B)
\begin{equation}\label{eq:noisepmb}
\bar{\kappa}_{\pm 1}^b=\pm i\eta\sqrt{\frac{\gamma}{2}}\frac{\Omega_1\cos\theta_L}{\gamma/2\mp i(\Delta_1\pm \nu')} ,
\end{equation}
so that $\kappa_{\pm1}^{b}=|\bar{\kappa}_{\pm 1}^b|^2$. The formal solution of Eqs.~(\ref{qlefin1})-(\ref{qlefin3}) at time $t=T_1$ reads
\begin{widetext}
\begin{eqnarray}
  \tilde{a}^{\dagger}( T_1) &= &g_{-1}(T_1)\tilde{a}^{\dagger}(0)+\chi_1^*f_1(T_1)\tilde{b}(0)
    +\chi^*_1\int^{T_1}_{0}{\rm d}s\; f_1(T_1-s)\left[\sqrt{2\kappa_b}\tilde{b}^{in}(s)
+\sqrt{2}\bar{\kappa}_{+1}^{b}\tilde{a}_{+}^{in}(s)-\sqrt{2}\bar{\kappa}_{-1}^{b}\tilde{a}_{-}^{in\dagger}(s)\right]
\nonumber \\
&& +\int^{T_1}_{0}{\rm d}s\;
g_{-1}(T_1-s)\left[\sqrt{2\kappa}\tilde{a}^{in\dagger}(s)
+\sqrt{2}\bar{\kappa}_{L}\tilde{a}_{-}^{in\dagger}(s)\right], \label{asol1}\\
  \tilde{b}( T_1)& = &\chi_1 f_1(T_1)\tilde{a}^{\dagger}(0)+ g_{+1}(T_1)\tilde{b}(0)
    +\int^{T_1}_{0}{\rm d}s\; g_{+1}(T_1-s)\left[\sqrt{2\kappa_b}\tilde{b}^{in}(s)
+\sqrt{2}\bar{\kappa}_{+1}^{b}\tilde{a}_{+}^{in}(s)-\sqrt{2}\bar{\kappa}_{-1}^{b}\tilde{a}_{-}^{in\dagger}(s)\right]\nonumber
\\
&& +\chi_1 \int^{T_1}_{0}{\rm d}s\;
f_1(T_1-s)\left[\sqrt{2\kappa}\tilde{a}^{in\dagger}(s)
+\sqrt{2}\bar{\kappa}_{L}\tilde{a}_{-}^{in\dagger}(s)\right],
\label{bsol1} \end{eqnarray} \end{widetext} where we have
introduced the time-dependent functions\begin{eqnarray}
  g_{\pm 1}(t) &=& e^{-\kappa_{1S}t}\left[\cosh(\theta_1t )\pm \frac{\kappa_{1D}}{\theta_1}\sinh(\theta_1t )\right], \label{gpm1}\\
  f_1(t) &=& e^{-\kappa_{1S}t}\frac{1}{\theta_1}\sinh(\theta_1t),
\end{eqnarray} and the parameters\begin{eqnarray}
  \kappa_{1S} &=& \frac{\kappa+\kappa_L+\kappa_b+\kappa_{+1}^{b}-\kappa_{-1}^{b}}{2}, \\
  \kappa_{1D} &=& \frac{\kappa+\kappa_L-\kappa_b-\kappa_{+1}^{b}+\kappa_{-1}^{b}}{2}, \\
  \theta_{1} &=& \sqrt{|\chi_1|^2+\kappa_{1D}^2}.
\end{eqnarray}
Equations~(\ref{asol1})-(\ref{bsol1}) describe the dynamics of the coupled motion and
cavity mode during the first pulse in presence of quantum noise. By setting all noise
terms to zero, $\kappa,\kappa_j=0$, they reproduce the well-known coherent two-mode
squeezing dynamics~\cite{milwal}, where entanglement monotonically increases as a
function of the interaction time $T_1$. The presence of quantum noise sets a limit to the
establishing of these dynamics.

\subsubsection{Dynamics between the two laser pulses}

In the time interval $T_1 \leq t \leq T$ the laser is turned off and consequently there
are no resonant photon scattering processes which couple the cavity mode and the atom's
vibrational motion. The HLE describing the system without laser excitation can be
obtained immediately from Eqs.~(\ref{qlefin1})-(\ref{qlefin3}) by setting $\Omega_1=0$
and thus $\chi_1=\bar{\kappa}_{+1}^{b}=\bar{\kappa}_{-1}^{b}=0$. The resulting HLE are
given by
\begin{eqnarray} && \dot{\tilde{a}}(t)=-\left(\kappa +\kappa_{L}\right)
\tilde{a}(t)+\sqrt{2\kappa}\tilde{a}^{in}(t)
+\sqrt{2}\bar{\kappa}_{L}\tilde{a}_{-}^{in}(t), \nonumber \\
&& \label{qlefin1b}\\
&& \dot{\tilde{b}}(t)= i\delta_1^b\tilde{b}(t)-\kappa_b
\tilde{b}(t)+\sqrt{2\kappa_b}\tilde{b}^{in}(t), \label{qlefin3b}
\end{eqnarray}
where off-resonant coupling between motion and
cavity mode is neglected, and $\delta_1^b$ is defined in
App.~\ref{App:A}. This latter term is due to the definition of the
reference frame rotating at frequency $\nu'$, which compensates
the laser-induced a.c.-Stark shift on the motion: In absence of
the laser, this component of the frequency $\nu'$ is unbalanced.
The assumption of neglecting off-resonant coupling between motion
and cavity mode is justified when the cavity-mode wavevector is
orthogonal to the axis of the motion, and thus there is no
mechanical coupling. In general, it is valid at zero order in the
expansion in the small parameter $\eta
g\cos\theta_c/|\omega_c-\omega_0|$, which hence imposes a
condition on the integration time $T$ under which
Eqs.~(\ref{qlefin1b})-(\ref{qlefin3b}) are valid. The solutions of
Eqs.~(\ref{qlefin1b})-(\ref{qlefin3b}) are \begin{widetext}
\begin{eqnarray}
  \tilde{a}(T) &=& e^{-\left(\kappa +\kappa_{L}\right)\left(T-T_1\right)}\tilde{a}(T_1)
  +\int_{T_1}^{T}ds e^{-\left(\kappa +\kappa_{L}\right)\left(T-s\right)}\left[\sqrt{2\kappa}\tilde{a}^{in}(s)
+\sqrt{2}\bar{\kappa}_{L}^*\tilde{a}_{-}^{in}(s)\right],\label{qlesol1b}\\
  \tilde{b}(T) &=& e^{(i\delta_1^b-\kappa_b)\left(T-T_1\right)}\tilde{b}(T_1)
  +\sqrt{2\kappa_b}\int_{T_1}^{T}ds
  e^{(i\delta_1^b-\kappa_b)\left(T-s\right)}\tilde{b}^{in}(s) \label{qlesol2b},
\end{eqnarray} \end{widetext} where $\tilde{a}(T_1)$ and
$\tilde{b}(T_1)$ are given by Eqs.~(\ref{asol1})-(\ref{bsol1}). They yield the observables of interest at time $T$, before the second laser
pulse is switched on.

\subsubsection{Dynamics during the second laser pulse}

We now consider that in the time interval $T \leq t \leq T_2$ a
square pulse of constant Rabi frequency $\Omega(t)=\Omega_2$ and
central angular frequency $\omega_{L2}$ illuminates the atom. We
denote by $\delta_2=\omega_{L2}-\omega_c$, and
$\Delta_2=\omega_{L2}-\omega_0$ the detuning of the laser frequency
from cavity mode and dipole transition, respectively. In the limit
in which processes where absorption of a laser photon and of a
phonon is resonant with emission of a cavity photon,
$\omega_{L2}\approx \omega_c-\nu$, we derive the HLE from
Eqs.~(\ref{priqle})-(\ref{lastqle}), describing the interaction
between the cavity mode and the vibrational motion during the second
laser pulse, in the presence of losses and noise processes due to
spontaneous emission, cavity decay, and vibrational heating,
\begin{widetext} \begin{eqnarray} &&
\dot{\tilde{a}}(t)=\chi_2\tilde{b}(t) -\left(\kappa
+\kappa_{L}\right) \tilde{a}(t)+\sqrt{2\kappa}\tilde{a}^{in}(t)
+\sqrt{2}\bar{\kappa}^*_{L}\tilde{a}_{+}^{in}(t)~, \label{qlefin12} \\
&& \dot{\tilde{b}}(t)=-\chi_2^* \tilde{a}(t) -\left(\kappa_b+\kappa_{+2}^{b}-\kappa_{-2}^{b}\right)
\tilde{b}(t)+\sqrt{2\kappa_b}\tilde{b}^{in}(t)
+\sqrt{2}\bar{\kappa}_{+2}^{b}\tilde{a}_{+}^{in}(t)-\sqrt{2}\bar{\kappa}_{-2}^{b}\tilde{a}_{-}^{in\dagger}(t)~, \label{qlefinb2} \end{eqnarray}
\end{widetext}
where parameters $\chi_2$ and $\kappa^b_{\pm 2}$ are found from $\chi_1$ and
$\kappa^b_{\pm 1}$ by replacing $\Omega_1 \to \Omega_2$, $\delta_1 \to \delta_2$, and
$\Delta_1 \to \Delta_2$ in Eqs.~(\ref{Chi:1:app}) and~(\ref{los1}). The details of the
derivation are reported in App.~\ref{App:A}.

The solutions of Eqs.~(\ref{qlefin12})-(\ref{qlefinb2}) at the end
of the second pulse, $t=T_2$, read \begin{widetext}
\begin{eqnarray}
  \tilde{a}( T_2) &= &g_{-2}(T_2-T)\tilde{a}(T)+\chi_2 f_2(T_2-T)\tilde{b}(T)
    +\chi_2\int^{T_2}_{T}ds\; f_2(T_2-s)\left[\sqrt{2\kappa_b}\tilde{b}^{in}(s)
+\sqrt{2}\bar{\kappa}_{+2}^{b}\tilde{a}_{+}^{in}(s)-\sqrt{2}\bar{\kappa}_{-2}^{b}\tilde{a}_{-}^{in\dagger}(s)\right]
\nonumber \\
&& +\int^{T_2}_{T}ds\;
g_{-2}(T_2-s)\left[\sqrt{2\kappa}\tilde{a}^{in}(s)
+\sqrt{2}\bar{\kappa}_{L}^*\tilde{a}_{+}^{in}(s)\right], \label{asol2}\\
  \tilde{b}( T_2)& = &-\chi_2^* f_2(T_2-T)\tilde{a}(T)+ g_{+2}(T_2-T)\tilde{b}(T)
    +\int^{T_2}_{T}ds\; g_{+2}(T_2-s)\left[\sqrt{2\kappa_b}\tilde{b}^{in}(s)
+\sqrt{2}\bar{\kappa}_{+2}^{b}\tilde{a}_{+}^{in}(s)-\sqrt{2}\bar{\kappa}_{-2}^{b}\tilde{a}_{-}^{in\dagger}(s)\right]\nonumber
\\
&& -\chi_2^* \int^{T_2}_{T}ds\; f_2(T_2-s)\left[\sqrt{2\kappa}\tilde{a}^{in}(s) +\sqrt{2}\bar{\kappa}_{L}^*\tilde{a}_{+}^{in}(s)\right],
\label{bsol2}
\end{eqnarray}
\end{widetext}
where we have introduced \begin{eqnarray}
  g_{\pm 2}(t) &=& e^{-\kappa_{2S}t}\left[\cos(\theta_2 t)\pm \frac{\kappa_{2D}}{\theta_2}\sin(\theta_2 t)\right], \label{gpm2}\\
  f_2(t) &=& e^{-\kappa_{2S}t}\frac{1}{\theta_2}\sin(\theta_2
 t),\label{f2} \end{eqnarray} and \begin{eqnarray}
  \kappa_{2S} &=& \frac{\kappa+\kappa_L+\kappa_b+\kappa_{+2}^{b}-\kappa_{-2}^{b}}{2}, \\
  \kappa_{2D} &=& \frac{\kappa+\kappa_L-\kappa_b-\kappa_{+2}^{b}+\kappa_{-2}^{b}}{2}, \\
  \theta_{2} &=& \sqrt{|\chi_2|^2-\kappa_{2D}^2}.
\end{eqnarray}
Equations~(\ref{asol2})-(\ref{bsol2}) give the cavity mode and motion at the end of the
second pulse. Setting all decay and noise sources to zero, we recover the ideal polariton
dynamics during the second pulse, namely a periodic dynamics at frequency $|\chi_2|$.
Ideally, then, the states of the motion and of the cavity mode are swapped when
$T_2-T=\delta T_2^0$, such that $|\chi_2|\delta T_2^0=(2\ell+1)\pi/2$, with $\ell$
integer number. At these values the function $f_2$ reaches its maximum, $f_2(\delta
T_2^0)|_{\kappa:j=0}=1$, while $g_{-2}$ vanishes, $g_{-2}(\delta T_2^0)|_{\kappa:j=0}=0$.
The effect of decay and noise is to damp the oscillators, hence to modify the oscillation
frequency and the behaviour of functions $f_2$, $g_{-2}$. In particular, the maximum
value of $f_2(t)$ is always smaller than unity, and the maxima of $f_2$ do not occur at
the same instants of time in which $g_{-2}(t)$ vanishes. We optimize the process by
setting $T_2-T$ such that it fulfills the condition $g_{-2}(T_2-T)=0$. Denoting by
$\delta T_2^{opt}= T_2-T$ the time interval fulfilling this condition, it satisfies the
relation
\begin{equation}
\delta T_2^{opt}\equiv \frac{1}{\theta_2}\arctan\frac{\theta_2}{\kappa_{2D}}~.
\label{condiT2}
\end{equation}

\section{Quantifying the entanglement between the two pulses}
\label{Sec:Correlation:Matrix}

The two time-separated pulses at the cavity output can be considered as two independent modes,
even if they originate from the same intracavity field at different times. In fact,
the output field $ a^{out}(t)$ is related to the intracavity field
$\tilde{a}(t)$ by the input-output relation \cite{milwal}
\begin{equation}\label{inout}
    \tilde{a}^{out}(t) = \sqrt{2 \kappa} \tilde{a}(t)- \tilde{a}^{in}(t)
\end{equation} and it is characterized by the commutation relation
$\left[\tilde{a}^{out}(t),\tilde{a}^{out}(t')^{\dagger}\right]=\delta(t-t')$.
In order to have a quantity directly related to the detected
field, we define the following integrated output field over a
generic measurement time $T_m$~\cite{vanEnk02}, \begin{equation}
    \tilde{a}^{out}_I(t,T_m)=\frac{1}{\sqrt{T_m}}\int^{t+T_m}_{t}dt'\;\tilde{a}^{out}(t').\label{integout}
\end{equation}
The field operators $\tilde{a}^{out}_I(t,T_m)$, form a class of
dimensionless bosonic operators,
$[\tilde{a}^{out}_I(t,T_m),\tilde{a}^{out \;\dagger}_I(t,T_m)]=1$,
and they commute, i.e., they describe independent modes, as soon as
they do not temporally overlap, that is,
$[\tilde{a}^{out}_I(t,T_m),\tilde{a}^{out \;\dagger}_I(t',T_m)]=0$
whenever $|t-t'| >T_m$. The two output pulses we are interested in
are therefore those associated with the operators
$\tilde{a}^{out}_I(T_1,T_m)$ and $\tilde{a}^{out}_I(T_2,T_m)$. In
the present scheme the two pulses are temporally separated and
therefore it is natural to consider $T_m < T-T_1$, which
automatically warrants the independence of the two integrated output
modes.

In order to characterize the entanglement between the two pulses,
one usually considers the amplitude and phase quadratures of the two
independent modes, which are in this case
\begin{eqnarray}
  X^{out}(T_j,T_m) &=& \frac{\tilde{a}^{out}( T_j,T_m)+ \tilde{a}^{out}( T_j,T_m)^{\dagger}}{\sqrt{2}},\\
  P^{out}(T_j,T_m) &=& \frac{\tilde{a}^{out}( T_j,T_m)- \tilde{a}^{out}(T_j,T_m)^{\dagger}}{i\sqrt{2}},
\end{eqnarray}
and construct the correlation matrix
\begin{equation}\label{correvout}
    V^{out}_{kl}=\frac{\left\langle \xi^{out}_k \xi^{out}_l+\xi^{out}_l \xi^{out}_k\right\rangle
    }{2},
\end{equation}
where we have defined the four-dimensional vector \begin{widetext}
$$ \xi^{out,T} =\left\{X^{out}(T_1,T_m),
P^{out}(T_1,T_m),X^{out}(T_2,T_m),P^{out}(T_2,T_m)\right\}. $$
\end{widetext}
In Eq.~(\ref{correvout}) the averaging corresponds to taking expectation values
with respect to the initial state of the system \emph{and} the
environment. We now proceed in determining its elements.

We use the definition in Eq.~(\ref{integout}), the input-output relation of
Eq.~(\ref{inout}), and the explicit solution for $\tilde{a}(t)$ in
the two relevant time intervals, $T_j \leq t \leq T_j+T_m$, $j=1,2$,
which is given by Eq.~(\ref{qlesol1b}), and get the following
expression of the integrated output fields as a function of the
intracavity fields at the end of the pulses $\tilde{a}(T_j)$ and of
the input noises:
\begin{equation}
a^{out}_I(T_j,T_m)=\alpha(T_m)\tilde{a}(T_J) + n^{in}(T_j,T_m),
\label{a:out:I}
\end{equation}
with the factor
\begin{equation}\label{alfa}
    \alpha(T_m)=
    \sqrt{\frac{2\kappa}{T_m}}\frac{\left[1-e^{-(\kappa+\kappa_L)T_m}\right]}{\kappa+\kappa_L},
\end{equation}
and the input noise term
\begin{widetext}
\begin{eqnarray}
&&n^{in}(T_j,T_m)=\int^{T_J+T_m}_{T_J}dt\;\frac{\tilde{a}^{in}(t)}{\sqrt{
T_m}}\left[\frac{\kappa-\kappa_L}{\kappa+\kappa_L}
-\frac{2\kappa}{\kappa+\kappa_L}e^{-(\kappa+\kappa_L)(T_J+T_m-t)}\right]
\nonumber \\
&&+\bar{\kappa}_L\sqrt{\frac{4\kappa}
{T_m}}\int^{T_J+T_m}_{T_J}dt\;\tilde{a}_{-}^{in}(t)\frac{\left[1-e^{-(\kappa+\kappa_L)(T_J+T_m-t)}\right]}{\left(\kappa+\kappa_L\right)}.
\end{eqnarray} \end{widetext}
Using Eq.~(\ref{a:out:I}) in Eq.~(\ref{correvout}), we find that the correlation matrix
$V^{out}$ can be decomposed into the sum of three contributions,
\begin{equation}\label{vout} V^{out}=\alpha(T_m)^2
V+V^{in}+V^{mix}, \end{equation} where $V^{in}$ is the contribution due to the input
noise term $n^{in}(T_j,T_m)$, $V^{mix}$ is the contribution due to the correlation
between the intracavity fields at the end of the pulses $\tilde{a}(T_J)$ and the input
noises, and $V$ is the correlation matrix for the quadratures of the intracavity fields
$X(T_j) =(\tilde{a}( T_j)+ \tilde{a}( T_j)^{\dagger})/\sqrt{2}$, $ P(T_j) = -i(\tilde{a}(
T_j)- \tilde{a}(T_j)^{\dagger})/\sqrt{2}$. Its elements have the form
\begin{equation}
\label{correv}
    V_{kl}=\frac{\left\langle \xi_k \xi_l+\xi_l \xi_k\right\rangle
    }{2},
\end{equation}
where we have defined the four-dimensional vector $\xi^{T}
=(X(T_1),P(T_1),X(T_2),P(T_2))$, and are reported in
App.~\ref{App:B}. Using the correlation functions of the input
noises $\tilde{a}^{in}(t)$ and $\tilde{a}_{-}^{in}(t)$ one finds
that $V^{in}$ is proportional to the $4 \times 4$ identity matrix
\begin{equation}\label{vin}
V^{in}=\frac{1}{2}\left[1-\alpha(T_m)^2\right]\delta_{ij},
\end{equation} while $V^{mix}$ has only four nonzero terms, which
are all identical,
$V^{mix}_{13}=V^{mix}_{31}=V^{mix}_{24}=V^{mix}_{42}={\cal
V}^{mix}$, with
\begin{widetext} \begin{equation}
    {\cal V}^{mix}=\alpha(T_m)
    \sqrt{\frac{\kappa}{2T_m}}\frac{e^{-(\kappa+\kappa_L)(T-T_1)}g_{-2}(T_2-T)}{\kappa+\kappa_L}
    \left\{e^{-(\kappa+\kappa_L)T_m}-1\right\}. \label{V:mix} \end{equation} \end{widetext}
Equation~(\ref{V:mix}) shows that the contributions due to
correlations between intracavity fields at the end of the pulse and
input noise are zero as soon as we choose the optimal transfer
condition $g_{-2}(T_2-T)=0$ of Eq.~(\ref{condiT2}) (They are in any
case negligible when $(\kappa+\kappa_L)(T-T_1)\gg 1$).

Let us now consider what is the optimal integration time $T_m$, such that $V^{out}\approx
V$, i.e.\, the correlation matrix at the cavity output reproduces the correlation matrix
between the intracavity fields. From Eqs.~(\ref{vout}) and (\ref{vin}) we see that $T_m$
must be chosen such that the quantity $\alpha(T_m)$ is as close as possible to unity. In
this case, most of the intracavity field is detected at the cavity output at the end of
the pulse, and at the same time the contribution of the input noise is negligible. From
Eq.~(\ref{alfa}) one gets that $\alpha(T_m)\le \alpha_{max}$, where $\alpha_{max}\simeq
0.9/\sqrt{1+\kappa_L/\kappa}<1$, and it is achieved for $T_m^0 \simeq
1.25/(\kappa+\kappa_L)$.

In order to establish the conditions under which the two output
pulses are entangled we consider the logarithmic negativity
$E_{\mathcal{N}}$, a quantity which has been already proposed as a
measure of entanglement \cite{werner}. In the continuous variable
case $E_{\mathcal{N}}$ can be defined as \cite{Salerno1}
\begin{equation}
E_{\mathcal{N}}=\max [0,-\ln 2\eta ^{-}], \label{logneg}
\end{equation}
where
\begin{equation}
\eta ^{-}\equiv 2^{-1/2}\left[ \Sigma (V^{out})- \left[ \Sigma
(V^{out})^{2}-4\det V^{out}\right] ^{1/2}\right] ^{1/2},
\label{Sympl_eigenv}
\end{equation}
with $\Sigma (V^{out})\equiv \det A+\det B-2\det C$, and we have used the $2\times2$
block form of the correlation matrix
\begin{equation}
V^{out}\equiv \left(
\begin{array}{cc}
A & C \\
C^{T} & B
\end{array}
\right) . \label{blocks}
\end{equation}
Therefore, a Gaussian state is entangled if and only if $ \eta ^{-}<1/2$. This is
equivalent to Simon's necessary and sufficient entanglement criterion for Gaussian states
of a non-positive partial transpose \cite{simon}, which can be written as $4\det V <
\Sigma -1/4$.

We finally comment on our choice to quantify the entanglement in terms of the logarithmic
negativity instead of EPR variances~\cite{Reid}. The latter would seem a natural choice,
but in fact they provide an unambiguous characterization of entanglement {\it only} for
simple examples of CV two-mode entangled states, such as the two-mode squeezed state. Any
reasonable entanglement measure has to be invariant under local transformations of the
quadratures of each mode separately. In our system, however, the common definition of EPR
correlations~\cite{Salerno1,simon} reads
\begin{eqnarray} &&\xi_{EPR}=
\frac{1}{2}\left[\Delta(X^{out}(T_1,T_m)-X^{out}(T_2,T_m))^2
\right. \nonumber
\\
&&\left. +\Delta(P^{out}(T_1,T_m)+P^{out}(T_2,T_m))^2\right],
\nonumber \end{eqnarray}
with $\Delta(A)^2$ the variance of $A$, and depends upon the chosen set of quadratures,
i.e.\ it does not possess such invariance. The light pulses at the cavity output are in a
two-mode squeezed state only in the ideal limit when the noise contributions to the
dynamics are negligible, and when the ion's motional state created by the first pulse is
perfectly transferred to the second pulse. Under realistic conditions the state of the
two pulses is rather different from a two-mode squeezed state; therefore using EPR
variances would give an ambiguous quantification of the generated entanglement.
Nonetheless, there is a connection between squeezing and the measure of entanglement
provided by the logarithmic negativity. To be more specific, a simple and direct
quantitative connection between EPR variances and $E_{\mathcal{N}}$ can be found in the
case of \emph{symmetric} bipartite states, i.e., states which are invariant under
exchange of the two modes (the two pulses in our case) \cite{Salerno1}. In such a case,
the quantity $2 \eta^- = \exp(-E_{\mathcal{N}})$ gives the largest amount of EPR
correlations, that is, the minimum achievable value of $\xi_{EPR}$ which can be attained
in the CV bipartite state by means of local operations, i.e. by considering all possible
linear combinations of the quadratures of each pulse.

\section{Results}
\label{Sec:Results}

We now analyze the basic requirements and the efficiency of this scheme using parameters
accessible in present experiments with atoms in resonators. Our considerations follow and
extend the corresponding discussion in~\cite{Morigi06}. For comparison, we will discuss
along with some realistic sets of parameters an idealised case where the cavity decay
rate is set to a very small value, and the noise terms are suppressed.

The atom's internal degrees of freedom need to provide an optical dipole transition which
couples to both the laser and the cavity mode. A suitable example would be an ${F}\!=0
\leftrightarrow {F^{\prime}}\!=\!1$ closed atomic transition with the quantization axis
$\vec{B}$ along the cavity axis, and $\vec{B}$, $\vec{k}_L$, and laser polarization
$\vec{E}_L$ mutually orthogonal. Ideal candidates would then be alkali-earth-metal atoms
or alkali-earth-metal-like ions, but other geometries can be found which allow for the
realization of this scheme using also alkali atoms or alkali-like ions. We consider an
alkali-earth atom in the geometrical configuration $\theta_L=0$, $\theta_c=\pi/2$, and
$\phi_c=0$. This means that the trap center coincides with an antinode of the cavity
mode, and that the motion takes place along the direction of the laser beam and
orthogonally to the cavity axis, such that there is no mechanical effect of the cavity
field on the motion. This assumption is not strictly necessary but it is made here in
order to simplify the discussion.

In order to favor motional Raman transitions over resonant scattering, we assume a value
of the atom-field detuning $\Delta$ which exceeds the values of the coupling strengths
$\Omega$ and $g_c\sqrt{n}$, ($n$ is the average number of cavity photons), and of the
atomic transition linewidth $\gamma$. Taking a typical value of the linewidth of an
optical dipole transition, $\gamma=2\pi\times 5$~MHz, we choose $|\Delta|=2\pi\times 120$
MHz, $\Omega=2\pi\times 10$~MHz and $g_c=2\pi\times 1$~MHz, which are accessible values
for state-of-the-art experiments with trapped atoms or ions in resonators
\cite{Keller04,Mundt02,Maurer04}.

We also consider the motion to be restricted to the Lamb-Dicke regime, with a Lamb-Dicke
parameter $\eta=0.1$. This leads to $|\chi_1|,|\chi_2| \simeq 2\pi\times 8$~kHz for the
coupling constants, while the loss rates associated with the various scattering processes
are $\kappa_L \sim \kappa^b_{\pm j} \simeq 2\pi\times 0.2$~kHz. The trap frequency can be
set to $\nu \simeq 2\pi\times 1$ MHz, which is typical in ion trap experiments
\cite{Eschner03}. The heating
rate of the vibrational motion may be estimated as $\kappa_h \simeq 2\pi\times 20$~Hz
\cite{Lucas2007,Labaziewicz2007}.

The value of the cavity decay rate $\kappa$ must warrant the coherent creation of
correlations during the pulses ($\kappa T_{1,2} \ll 1$), as well as spectral resolution
of the sidebands ($\kappa \ll \nu$). In the examples we discuss we assume values that
range from $\kappa \simeq 2\pi\times 20$~kHz, which is experimentally accessible
\cite{Maurer04,Sauer04}, down to $2\pi\times 1$~kHz, which is more difficult to
reach with present-day technology, but serves as an idealised case for comparison.

Given the parameters, finally the laser pulse durations $T_1$ and $T_2-T$, as well as
their separation $T$ must be adjusted in order to (i) create significant entanglement
between the first pulse and the motion, and (ii) efficiently realise the quantum state
transfer between the motion and the second pulses. At the same time the motion, which
acts as intermediate quantum memory, needs to remain coherent during laser excitation and
cavity output.

\begin{center}
\begin{figure}[htb]
\includegraphics[width=0.45\textwidth]{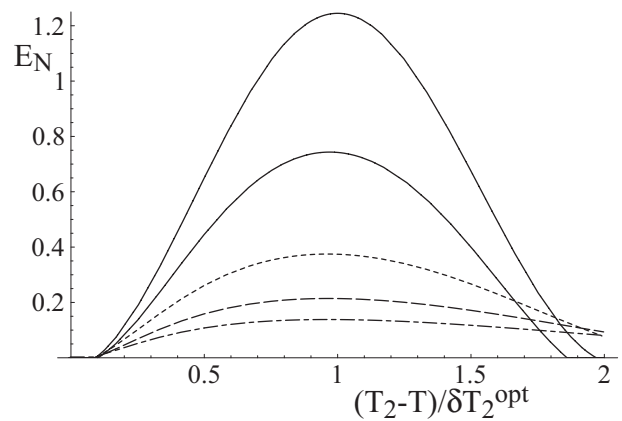}
\caption{Logarithmic negativity $E_{\mathcal{N}}$ versus normalized duration of the
second pulse $(T_2-T)/\delta T_2^{opt}$ (see Eq.~(\protect\ref{condiT2})) at four
different values of the cavity decay rate. From bottom to top, $\kappa =2\pi\times (16,
11, 6.4, 0.8, 0.8)$~kHz; the top solid curve is plotted for comparison and corresponds to
the smallest value of $\kappa$ with all other noise terms set to zero. The other
parameter values are $(|\Delta|, \gamma, \Omega, g_c, \nu) = 2\pi\times (120, 5, 10, 1,
1)$~MHz and $\eta =0.1$, yielding $(|\chi_{1}|, |\chi_{2}|, \kappa_L, \kappa^b_{\pm j})
=2\pi\times (8.5, 8.5, 0.18, 0.18)$~kHz; the heating rate is $\kappa_h =2\pi\times
20$~Hz. The duration of the first pulse is fixed at $T_1 = 40~\mu$s $\sim
2/|\chi_1|^{-1}$, while for each curve, the time interval between the two pulses is
related to the cavity decay time according to $T-T_1 = 2/(\kappa+\kappa_L) = (20, 28, 49,
327, 400)~\mu$s (bottom to top).} \label{fig:logneg1}
\end{figure}
\end{center}

We first focus on the dependence upon the duration of the second pulse, $T_2-T$. The
optimization of the state transfer from the center-of-mass motion to the cavity mode
discussed in Sec. IIB.3 yielded the condition of Eq.~(\ref{condiT2}), $T_2-T=\delta
T_2^{opt}$. Therefore we expect the entanglement to be maximum around this condition.
This is confirmed by Fig.~\ref{fig:logneg1}, where the logarithmic negativity is plotted
versus the normalized duration of the second pulse, $(T_2-T)/\delta T_2^{opt}$ at various
values of the cavity decay rate. The duration of the first pulse has been fixed at $T_1 =
40 $ $\mu$sec $\simeq 2/|\chi_1|^{-1}$, while the time interval between the two pulses
has been chosen to be related to the cavity decay time according to $T-T_1 =
2/(\kappa+\kappa_L)$; the other parameter values are those discussed above. It is evident
that the entanglement between the pulses is optimized when the duration of the second
pulse satisfies Eq.~(\ref{condiT2}).

\begin{center}
\begin{figure}[htb]
\includegraphics[width=0.45\textwidth]{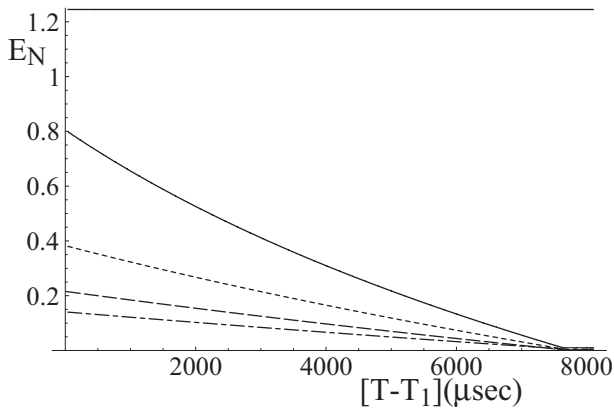}
\caption{Logarithmic negativity $E_{\mathcal{N}}$ versus the time separation between the
two pulses $T-T_1$ at the same values of the cavity decay rate $\kappa$ as in
Fig.~\protect\ref{fig:logneg1}. The other parameter values are also the same as in
Fig.~\protect\ref{fig:logneg1} except that the duration of the second pulse has been
fixed at the optimal value $T_2-T=\delta T_2^{opt} = (19, 22, 24, 29, 29)~\mu$s (bottom
to top).} \label{fig:logneg2}
\end{figure}
\end{center}

The dependence of the logarithmic negativity $E_{\mathcal{N}}$ upon the time separation
$T-T_1$ between the two pulses is shown in Fig.~\ref{fig:logneg2}. Parameter values are
the same as in Fig.~\ref{fig:logneg1} except that we have fixed the duration of the
second pulse at the optimal value $T_2-T=\delta T_2^{opt}$ given by Eq.~(\ref{condiT2}).
Entanglement (i.e.\ $E_{\mathcal{N}}$) decays essentially linearly, and notably its
lifetime is independent from the cavity decay rate $\kappa$. In fact, $E_{\mathcal{N}}$
always vanishes when $T-T_1 \simeq \kappa_h^{-1}$, i.e.\ the two pulses are entangled
provided that their separation is not larger than the vibrational heating time
$\kappa_h^{-1}$. This is not surprising because the vibrational motion acts as the
continuous-variable quantum memory mediating the entanglement, and thus the heating time
limits the achievable coherent storage time.

\begin{center}
\begin{figure}[htb]
\includegraphics[width=0.45\textwidth]{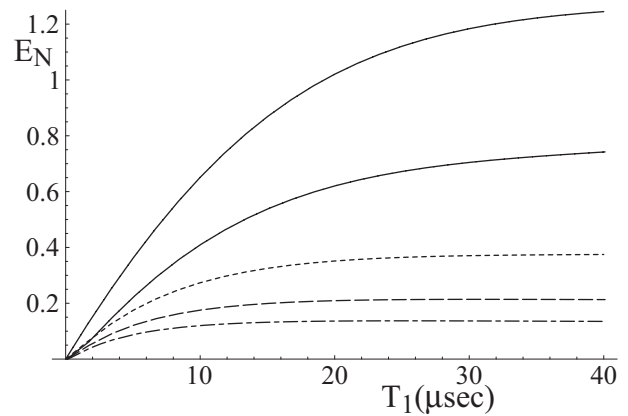}
\caption{Logarithmic negativity $E_{\mathcal{N}}$ versus duration of the first laser
pulse $T_1$, for the same values of $\kappa$ as in the previous figures. The time
interval between the two pulses and the duration of the second pulse are set for each
curve according to $T-T_1 = 2/(\kappa+\kappa_L)$ and $T_2-T=\delta T_2^{opt}$. The other
parameter values are the same as before. } \label{fig:logneg3}
\end{figure}
\end{center}

Finally, the dependence of the logarithmic negativity upon the duration of the first
pulse $T_1$ is shown in Fig.~\ref{fig:logneg3}. The other two timing parameters are, for
each curve, $T_2-T=\delta T_2^{opt}$ as in Fig.~\ref{fig:logneg2}, and $T-T_1 =
2/(\kappa+\kappa_L)$ as in Fig.~\ref{fig:logneg1}. The rest of the parameters have the
same values as before.
$E_{\mathcal{N}}$ is always increasing and then tends to saturate at a value that, as
expected, is larger for smaller cavity decay rates. In presence of noise one finds an
empirical expression for the asymptotic logarithmic negativity given by
$E_{\mathcal{N}}^{asym} \simeq \ln\left[4|\chi_1|/\kappa\right]/4$. This behavior can be
intuitively explained by the fact that the first laser pulse entangles the cavity mode
and the vibrational motion like in a parametric amplifier, and this continuous variable
entanglement increases for increasing $T_1$. Cavity losses, however, limit the
entanglement generation and  are ultimately responsible for the saturation of the
entanglement at larger $T_1$.

From Fig.~\ref{fig:logneg3} it can be seen that choosing a large $T_1$, in the saturation
regime, has two advantages: i) entanglement is maximized; ii) the scheme is insensitive
to fluctuations of $T_1$. In must be kept in mind, however, that we are considering the
entanglement between two output light pulses which are counted for a time interval
$T_m^0=1.25/(\kappa+\kappa_L)$ starting only when each exciting laser pulse has finished
(see Sec. III). This means that for large values of $T_1$ the detection of the two
entangled pulses is more difficult, especially when the cavity decay rate is large. In
these latter cases, the number of photons leaving the cavity \emph{during} the first
excitation pulse is larger than the average number of photons in the two detected output
pulses. In the saturation regime $T_1=40$ $\mu$sec, the average number of photons per
entangled pulse is $\bar{n}(T_1)=\left\langle \tilde{a}^{\dag}(T_1)
\tilde{a}(T_1)\right\rangle = (12.38, 3.65, 1.55, 0.76)$ for $\kappa =2\pi\times (0.8,
6.4, 11, 16)$~kHz, where we have set $T_m^0=1.25/(\kappa+\kappa_L)$, while the other
parameters are as in the figures.

\section{Conclusions} \label{Sec:Conclusions}

To conclude, we have characterized the quantum correlations of two temporally separated
entangled light pulses, emitted from a single atom inside an optical cavity, in the
set-up first proposed in~\cite{PRL, Morigi06}. By means of a quantum noise analysis we
have quantified the amount of entanglement one can extract from this system for
experimentally accessible parameter regimes, and we have shown that the quantum motion of
a single trapped particle is an efficient quantum medium which creates and mediates
entanglement on demand between subsequent radiation pulses.

This scheme offers promising perspectives for atom-photon interfaces and for devising new
cryptographic schemes exploiting time-correlated pulses and continuous alphabets, thus
extending those based on time-bin entangled photon pairs~\cite{Gisetal02}. The scheme may
also be easily generalised to the generation of two pulses of different frequencies, or
even at vastly different wavelengths, by varying appropriately the frequency or
wavelength of the second excitation pulse.

An interesting outlook is to study the scalability of the scheme when the number of atoms
composing the quantum medium is increased in a controlled way, hence characterizing which
resources the collective excitations of the medium may offer for creating entangled
light.

\section{Acknowledgements} The authors acknowledge discussions with
Stefano Mancini. This work was partly supported by the European Commission ("CONQUEST",
MRTN-CT-2003-505089; "EMALI", MRTN-CT-2006-035369; "SCALA", Contract No.\ 015714), by the
Spanish Ministerio de Educaci\'on y Ciencia ("QOIT", Consolider Ingenio 2010
CSD2006-00019; "LACSMY", FIS2004-05830; "QLIQS", FIS2005-08257; "QNLP", FIS2007-66944;
Ramon-y-Cajal program), and by the Italian Ministero dell Universit\`a e Ricerca
(PRIN-2005024254).

\begin{appendix}

\section{Mechanical effects in the Langevin force} \label{App:0}

We derive the Langevin force considering the mechanical effects of
the spontaneously emitted photon on the atom. At this purpose, we
consider a simple model, constituted by a dipole of frequency
$\omega_0$ at position $x$ and Hamiltonian of the center of mass
$H_{\rm CM}$, \begin{eqnarray*} H_{\rm at}=\hbar
\omega_0|e\rangle\langle e|+H_{\rm CM} \end{eqnarray*} The dipole
is coupled to the modes of the e.m.-field at frequency $\omega_s$
and Hamiltonian $$H_{\rm
emf}=\sum_s\hbar\omega_sa^{\dagger}_sa_s,$$ with interaction
Hamiltonian \begin{eqnarray*} H_{\rm int}=\hbar\sum_sg_s{\rm
e}^{ik_sx}\sigma^{\dagger}a_s+{\rm H.c.} \end{eqnarray*} where
$g_s$ is the vacuum Rabi frequency. The formal solution of the
Heisenberg equation for $a_s$ gives \begin{equation} a_s(t)={\rm
e}^{-{\rm i}\omega_st}a_s(0)-ig_s\int_0^td\tau{\rm
e}^{-i\omega\tau}{\rm e}^{-ik_sx(t-\tau)}\sigma(t-\tau)
\end{equation} We substitute this result into the Heisenberg
equation for $\sigma$, and obtain \begin{eqnarray}
\dot{\sigma}&=&-i\omega_0\sigma +i\sigma_z \sum_sg_s{\rm e}^{ik_sx}{\rm e}^{-i\omega_st}a_s(0) \\
& &+\sigma_z \sum_sg_s^2\int_0^t{\rm e}^{ik_s[x(t)-x(t-\tau)]}{\rm e}^{-i\omega_s\tau}\sigma(t-\tau)\nonumber \end{eqnarray} We make now the
Markov approximation, assuming that the characteristic frequencies of the center of mass motion are much smaller than the optical frequencies of
the e.m.-field, which couple quasi-resonantly with the dipole. Hence, in the integral we approximate $x(t-\tau)\approx x(t)$. Taking
$\sigma(t-\tau)\approx {\rm e}^{i\omega_0\tau}\sigma(t)$, we obtain \begin{eqnarray}
\dot{\sigma}&=&-i\omega_0\sigma +\sigma_z i\sum_sg_s{\rm e}^{ik_sx}{\rm e}^{-i\omega_st}a_s(0) \\
& &-\sigma(t) \sum_sg_s^2\int_0^\infty{\rm
e}^{i(\omega_0-\omega_s)\tau}{\rm d}\tau\nonumber \end{eqnarray} The integral
gives a real term, the linewidth
$\gamma=2\pi\sum_sg_s^2\delta(\omega_0-\omega_s)$, and an imaginary
part, the Cauchy principal value shifting the transition
frequency. Including this shift in the value of $\omega_0$ we
obtain \begin{eqnarray} \dot{\sigma}&=&-i\omega_0\sigma
-\frac{\gamma}{2}\sigma +i\sigma_zF_0(t) \end{eqnarray} with the
noise source \begin{equation} F_0(t)=\sum_sg_s{\rm e}^{ik_sx}{\rm
e}^{-i\omega_st}a_s(0) \end{equation} such that $\langle
F_0(t)F_0(t')^{\dagger}\rangle=\gamma\delta(t-t')$. Let us now
investigate the form of $F_0(t)$ when the center of mass motion is
a harmonic oscillator, $H_{\rm CM}=H_{\rm mec}$ given by
Eq.~(\ref{H:mec}). Applying the Lamb-Dicke expansion, $F_0(t)$
takes the form \begin{eqnarray}
F_0(t)&=&\sum_sg_s{\rm e}^{-i\omega_st}a_s(0)\\
& &\times
\left(1-\frac{\eta^2}{2}\cos^2\theta_s(2b^{\dagger}b+1)+i\eta\cos\theta_s(b^{\dagger}+b)\right)\nonumber
\end{eqnarray} with $k_sx=\eta\cos\theta_s(b^{\dagger}+b)$, where
$\theta_s$ is the angle between the wave vector of the emitted photon and the axis of the motion. Taking the continuous limit of the sum, we
separate the integrals over the modulus and the polar angle, and over the azimuthal angle $\theta=\theta_s$, obtaining $F_0(t)\approx F(t)$,
with
\begin{eqnarray}
F(t)&=&\sqrt{\gamma}\int {\rm d}\cos\theta \sqrt{{\cal N}(\cos\theta)} f_{\theta}^{in}(t)\\
& &\times \left(1-\frac{\eta^2}{2}\cos^2\theta(2b^{\dagger}b+1)+i\eta\cos\theta(b^{\dagger}+b)\right)\nonumber,
\end{eqnarray} where ${\cal N}(\cos\theta)$ is the dipole pattern
of emission, $\int {\rm d}\cos\theta {\cal N}(\cos\theta)=1$ and
$f_{\theta}^{in}(t)$ are the angle-dependent Langevin forces,
\begin{equation} \langle
f^{in}_{\theta}(t)f^{in\dagger}_{\theta'}(t')\rangle=\delta(\theta-\theta')\delta(t-t'),
\end{equation}
and which have zero mean value.

\section{Effective Quantum Langevin Equations} \label{App:A}

{\bf First laser pulse.} Starting from
Eqs.~(\ref{HLE:0:1})-(\ref{HLE:0:2}), which are defined in the
reference frame of the laser, we move to a frame rotating at the
effective vibrational angular frequency $\nu' \simeq \nu$, and we
neglect all the terms oscillating at $\nu'$ or larger. This
approximation is justified in the regime we consider, where we
assume that the Stokes and anti-Stokes sidebands are spectrally
resolved. The operators in this reference frame are connected to the
ones in the reference frame of the laser by the transformation
$\tilde{a}^{\dagger}(t)= e^{i\nu 't} a^{\dagger}(t)$, $\tilde{b}(t)=
e^{i\nu 't} b(t)$, and their equations of motion have the explicit
form \begin{widetext} \begin{eqnarray} && \label{QLE_01}
\dot{\tilde{a}}^{\dagger}(t)=i\left(\nu'-\delta_1 \right)
\tilde{a}^{\dagger}(t) -\kappa
\tilde{a}^{\dagger}(t)+\sqrt{2\kappa}\tilde{a}^{in}(t)^{\dagger}
-\int_0^t {\rm d}s e^{-\left(\frac{\gamma}{2}+i\Delta_1 \right)s}
\left[B^{\dagger}(t-s)+F^{\dagger}(t-s)\right]e^{i\nu 't} \nonumber \\
&& \times \left[g_c \cos \phi_c \left(1-\frac{\eta^2}{2}\cos^2
\theta_c (2 \tilde{b}^{\dagger}\tilde{b}+1)\right)-\eta g_c
\sin\phi_c \cos\theta_c\left(\tilde{b}(t)e^{-i\nu
't}+\tilde{b}^{\dagger}(t)e^{i\nu 't}\right)\right], \\
&& \dot{\tilde{b}}(t)=i\left(\nu'-\nu \right) \tilde{b}(t)
-\kappa_b \tilde{b}(t)+\sqrt{2\kappa_b}\tilde{b}^{in}(t) +\int_0^t
{\rm d}s e^{-\left(\frac{\gamma}{2}-i\Delta_1
\right)s} \left[B(t-s)+F(t-s)\right]e^{i\nu 't}\label{QLE_0N} \\
&& \times \Bigl[i\eta \Omega_1^* \cos \theta_L +\eta g_c^*
\sin\phi_c \cos\theta_c\tilde{a}^{\dagger}(t)e^{-i\nu 't} +
\eta^2\tilde{b}(t)e^{-i\nu 't}\left(\Omega_1^*\cos^2
\theta_L+g_c^* \cos\phi_c
\cos^2\theta_c\tilde{a}^{\dagger}(t)e^{-i\nu 't}\right)\nonumber\\
&&\hspace{2cm}+ {\rm i}\eta\sqrt{\gamma}\langle f^{in\dagger}_{\theta}(t)\cos\theta\rangle_{\theta}+\eta^2\sqrt{\gamma}\tilde{b}(t)e^{-i\nu
't}\langle f^{in\dagger}_{\theta}(t)\cos^2\theta \rangle_{\theta}\Bigr] -\int_0^t {\rm d}s e^{-\left(\frac{\gamma}{2}+i\Delta_1
\right)s} \left[B^{\dagger}(t-s)+F^{\dagger}(t-s)\right]e^{i\nu 't} \nonumber \\
&& \times \Bigl[-i\eta \Omega_1 \cos \theta_L +\eta g_c \sin\phi_c
\cos\theta_c\tilde{a}(t)e^{i\nu 't}+ \eta^2\tilde{b}(t)e^{-i\nu
't}\left(\Omega_1\cos^2 \theta_L+g_c\cos\phi_c
\cos^2\theta_c\tilde{a}(t)e^{i\nu 't}\right)\nonumber\\
&&\hspace{2cm}-{\rm i}\eta\sqrt{\gamma}\langle f^{in}_{\theta}(t)\cos\theta\rangle_{\theta}+\eta^2\sqrt{\gamma}\tilde{b}(t)e^{-i\nu 't}\langle
f^{in}_{\theta}(t)\cos^2\theta \rangle_{\theta} \Bigr]\nonumber, \end{eqnarray} \end{widetext} where $\langle \ldots\rangle_{\theta}$ in the
equation for $\dot{\tilde{b}}(t)$ denotes the average over the azimuthal angle $\theta$ with weight given by the dipole pattern of emission
${\cal N}(\cos\theta)$, and we have introduced the noise operators $\tilde{a}^{in}(t)\equiv e^{-i\nu 't} a^{in}(t)$ and $\tilde{b}^{in}(t)\equiv
e^{i\nu 't} b^{in}(t)$, which are still delta-correlated. We use the explicit expression for $B(t-s)$, thereby neglecting the terms oscillating
at $\nu '$ or faster, and perform the time integrals by making the Markovian approximation $\exp\{-(\gamma/2\pm i\Delta_1 +i m \nu')s\} \approx
\delta(s)/(\gamma/2\pm i\Delta_1 +i m \nu')$, for $m=-1,0,1$.

After long, but straightforward calculations we get the final,
effective HLE at leading order in the Lamb-Dicke parameter, which
read \begin{widetext} \begin{eqnarray} \label{QLE:a1}
\dot{\tilde{a}}^{\dagger}(t)&=&i\left(\nu'-\delta_1+\delta '
\right) \tilde{a}^{\dagger}(t) +\chi_1^*\tilde{b}(t) -\left(\kappa
+\kappa_{L}\right)
\tilde{a}^{\dagger}(t)+\sqrt{2\kappa}\tilde{a}^{in}(t)^{\dagger}
+\sqrt{2}\bar{\kappa}_{L}\tilde{a}_{-}^{in}(t)^{\dagger}+F_a, \\
\label{QLE:b} \dot{\tilde{b}}(t)&=&i\left(\nu'-\nu
-\delta_{1}^{b}\right) \tilde{b}(t) +\bar{\chi}_1
\tilde{a}^{\dagger}(t)
-\left(\kappa_b+\kappa_{+1}^{b}-\kappa_{-1}^{b}\right)
\tilde{b}(t)\\
& &+\sqrt{2\kappa_b}\tilde{b}^{in}(t)
+\sqrt{2}\bar{\kappa}_{+1}^{b}\tilde{a}_{+}^{in}(t)-
\sqrt{2}\bar{\kappa}_{-1}^{b}\tilde{a}_{-}^{in}(t)^{\dagger}+F_b~.\nonumber
\end{eqnarray} \end{widetext}

Here, the coefficients \begin{eqnarray} &&\chi_1=\eta \Omega_1
g_c^*\cos\phi_c\left(\frac{\cos\theta_L}{\Delta_1-\nu'+{\rm
i}\gamma/2} +\frac{{\rm
i}\tan\phi_c\cos\theta_c}{\Delta_1+{\rm i}\gamma/2}\right)~,\nonumber\\
&&\label{Chi:1}\\
&&\bar{\chi}_1=\eta \Omega_1
g_c^*\cos\phi_c\left(\frac{\cos\theta_L}{\Delta_1-\nu'-{\rm
i}\gamma/2} +\frac{{\rm
i}\tan\phi_c\cos\theta_c}{\Delta_1+{\rm i}\gamma/2}\right)~,\nonumber\\
&&\label{Chi:1_t} \end{eqnarray} correspond to the Raman processes in which laser photons are scattered into the cavity mode with a change in
the center-of-mass excitation. Since we assume $\gamma \ll |\Delta_1|$ (and also $\nu\ll|\Delta_1|$), we shall take $\bar{\chi}_1=\chi_1$, and
approximate $\chi_1$ with Eq.~(\ref{Chi:1:app}). The noise operators $\tilde{a}^{in}_{\pm}(t)$ are defined in Eqs.~(\ref{nos1})-(\ref{nos2}),
while the rates $\kappa_L$, $\kappa_{\pm 1}^b$ in Eqs.~(\ref{los1})-(\ref{Photon:scatter}). The noise scaling factors $\bar{\kappa}_{\pm 1}^b$
of Eq.~(\ref{eq:noisepmb} are generally given by \begin{eqnarray*}
&&\bar{\kappa}_{+1}^b=i\eta\sqrt{\frac{\gamma}{2}}\left(\frac{\Omega_1\cos\theta_L}{\gamma/2-i(\Delta_1+\nu')}
-\frac{\Omega_1^*\langle \cos\theta\rangle_{\theta}}{\gamma/2+i\Delta_1}\right)\\
&&\bar{\kappa}_{-1}^b=-i\eta\sqrt{\frac{\gamma}{2}}\left(\frac{\Omega_1\cos\theta_L}{\gamma/2+i(\Delta_1-\nu')} +\frac{\Omega_1^*\langle
\cos\theta\rangle_{\theta}}{\gamma/2-i\Delta_1}\right),
\end{eqnarray*} but they reduce to the expression of Eq.~(\ref{eq:noisepmb}) because the average over the dipole pattern gives $\langle
\cos\theta\rangle_{\theta}=0$.

The operators $F_a$ and $F_b$ in Eqs.~(\ref{QLE:a1})-(\ref{QLE:b})
represent non-linear noise terms, associated with incoherent
scattering processes. They give rise to a.c.-Stark shift and
losses, whose effect is in general detrimental for the
effectiveness of the two-mode squeezing processes, and are of the
form $F_a\sim \eta^2 g^2
\tilde{a}\tilde{b}^{\dagger}\tilde{b}/(\Delta+{\rm i}\gamma/2)$,
$F_b\sim \eta^2 g^2
\tilde{a}^{\dagger}\tilde{a}\tilde{b}/(\Delta+{\rm i}\gamma/2)$,
hermitian conjugates, and corresponding input noise operators.
These terms can be neglected in comparison with the laser induced
Raman scattering processes whenever the inequality
$\Omega_1\cos\theta_L\gg g_c\sqrt{n}\cos\theta_c$ is satisfied
($n$ is the average number of cavity photons), which is a
condition on the parameters and on the geometry of the setup. We
will assume this regime, and these terms will be neglected during
the laser pulses. In this regime, the frequency shifts of the
cavity mode and of the vibrational motion read \begin{eqnarray}
&&\delta'=\frac{(\Delta_1-\nu')|g_c|^2\cos^2\phi_c}{\gamma^2/4+(\Delta_1-\nu')^2} \label{eqfornu1} \\
&&\delta_{1}^{b}=2\eta^2 |\Omega_1|^2
\cos^2\theta_L\Delta_1\label{nu_b} \\
&&\left(\frac{ \gamma^2/4+\Delta_1^2-\nu'^2}{\left(\gamma^2/4+\Delta_1^2-\nu'^2\right)^2
+\nu'^2\gamma^2}-\frac{1}{\Delta_1^2+\gamma^2/4}\right)\nonumber.
\end{eqnarray}
These shifts must be taken into account when
tuning the frequency of the first laser pulse. In particular,
since the dynamics we seek relies on the resonant two-photon
processes, where a laser photon is absorbed and a the cavity
photon and a vibrational phonon are emitted, hence the cavity mode
frequency must be exactly at resonance with the Stokes sideband of
the driving laser, i.e., \begin{equation}
\omega_{L1}=\omega_c+\delta' +\nu+\delta\nu \label{res1} .
\end{equation} Here, $\delta\nu$ can be extracted from
Eq.~(\ref{nu_b}) when $|\delta\nu|\ll\nu$, which is satisfied
provided that $\eta |\Omega_1 | \ll |\Delta_1|$ and
$\eta|\Omega_1/\Delta_1|\ll\nu$, and reads
\begin{eqnarray}\label{renfreq}
  \delta\nu&\approx& \frac{2\Delta_1 \eta^2 |\Omega_1|^2
  \cos^2\theta_L\left(\gamma^2/4+\Delta_1^2-\nu^2\right)}{\left(\gamma^2/4+\Delta_1^2-\nu^2\right)^2+\nu^2\gamma^2}\nonumber\\
& &~-2\eta^2 |\Omega_1|^2 \cos^2\theta_L\frac{\Delta_1}{\Delta_1^2+\gamma^2/4}.\label{d:nu}
\end{eqnarray}
Consequently,
\begin{equation} \nu'=\nu+\delta\nu\end{equation} determines the
effective frequency of the Stokes sideband.

We notice that the resonant condition (\ref{res1}), giving the
relation $\omega_{L1}-\nu'=\omega_c+\delta'$, gives us a simple
relation between the observables in laboratory frame and the
"tilded" observables, which are connected by the relations
\begin{eqnarray} a_{\rm lab}(t)&=&
e^{-i(\omega_c+\delta')t}\tilde{a}(t)\\
b_{\rm lab}(t)&=& e^{-i\nu' t}\tilde{b}(t). \end{eqnarray}

{\bf Second laser pulse.} The main difference with the treatment
of the dynamics during the first pulse is that the second pulse is
set to a different resonance condition, e.g. \begin{equation}
\omega_{L2}= \omega_c-\nu^{\prime\prime} \label{Cond:2},
\end{equation} where $\nu^{\prime \prime}\simeq \nu$ and the
difference accounts for the a.c.-Stark shifts induced by the
second laser pulse on cavity and motion frequency. When
condition~(\ref{Cond:2}) is fulfilled, the cavity mode is resonant
with the anti-Stokes motional sideband of the laser light, with
angular frequency $\omega_{L2}+\nu^{\prime \prime}$. Spectral
resolution of this resonance is warranted when $\nu T_2\gg 1$ and
$\nu \gg \kappa,\kappa_b$.

In the reference frame rotating at the frequency of the cavity
mode, the HLE read \begin{widetext} \begin{eqnarray}
\label{QLE2nd1} && \dot{\tilde{a}}(t)=i\left(\nu^{\prime\prime}+\delta_2
\right) \tilde{a}(t)-\kappa
\tilde{a}(t)+\sqrt{2\kappa}\tilde{a}^{in}(t) +\int_0^t {\rm
d}s~e^{-\left(\frac{\gamma}{2}-i\Delta_2 \right)s}
\left[B(t-s)+F(t-s)\right]e^{i\nu ^{\prime\prime}t} \\
&& \times \left[-g_c^* \cos \phi_c \left(1-\frac{\eta^2}{2}\cos^2
\theta_c (2 \tilde{b}^{\dagger}\tilde{b}+1)\right) +\eta g_c^*
\sin\phi_c \cos\theta_c\left(\tilde{b}(t)e^{-i\nu^{\prime\prime}
t}+\tilde{b}^{\dagger}(t)e^{i\nu^{\prime\prime}t}\right)\right], \nonumber \\
&& \dot{\tilde{b}}(t)=i\left(\nu^{\prime\prime}-\nu \right) \tilde{b}(t)
-\kappa_b \tilde{b}(t)+\sqrt{2\kappa_b}\tilde{b}^{in}(t)+\int_0^t
{\rm d}s~e^{-\left(\frac{\gamma}{2}-i\Delta_2
\right)s}\left[B(t-s)+F(t-s)\right]e^{i\nu^{\prime\prime} t}\label{QLE2nd2} \\
&& \times \Bigl[i\eta \Omega_2^* \cos \theta_L +\eta g_c^*
\sin\phi_c \cos\theta_c\tilde{a}^{\dagger}(t)e^{i\nu^{\prime\prime}t} +
\eta^2\tilde{b}(t)e^{-i\nu^{\prime\prime}t}\left(\Omega_2^*\cos^2
\theta_L+g_c^* \cos\phi_c
\cos^2\theta_c\tilde{a}^{\dagger}(t)e^{i\nu^{\prime\prime}t}\right)\nonumber\\
&&\hspace{2cm}+ {\rm i}\eta\sqrt{\gamma}\langle f^{in\dagger}_{\theta}(t)\cos\theta\rangle_{\theta}+\eta^2\sqrt{\gamma}\tilde{b}(t)e^{-i\nu "
t}\langle f^{in\dagger}_{\theta}(t)\cos^2\theta \rangle_{\theta} \Bigr] -\int_0^t {\rm d}s~e^{-\left(\frac{\gamma}{2}+i\Delta_2
\right)s} \left[B^{\dagger}(t-s)+F^{\dagger}(t-s)\right]e^{i\nu^{\prime\prime}t} \nonumber \\
&& \times \Bigl[-i\eta \Omega_2 \cos \theta_L +\eta g_c \sin\phi_c
\cos\theta_c\tilde{a}(t)e^{-i\nu^{\prime\prime}t}+ \eta^2\tilde{b}(t)e^{-i\nu^{\prime\prime}
t}\left(\Omega_2\cos^2 \theta_L+g_c\cos\phi_c
\cos^2\theta_c\tilde{a}(t)e^{-i\nu^{\prime\prime}t}\right)\nonumber\\
&&\hspace{2cm}-{\rm i}\eta\sqrt{\gamma}\langle f^{in}_{\theta}(t)\cos\theta\rangle_{\theta}+\eta^2\sqrt{\gamma}\tilde{b}(t)e^{-i\nu" t}\langle
f^{in}_{\theta}(t)\cos^2\theta \rangle_{\theta} \Bigr]\nonumber \end{eqnarray} \end{widetext} where the noise operators $\tilde{a}^{in}(t)$ and
$\tilde{b}^{in}(t)$ are the same as in Eqs.~(\ref{QLE_01})-(\ref{QLE_0N}). Following the procedure outlined before, we finally get the coupled
HLE
\begin{widetext}
\begin{eqnarray} \label{QLE:a12nd}
\dot{\tilde{a}}(t)&=&i\left(\nu^{\prime}+\delta_2-\delta' \right)
\tilde{a}(t) +\chi_2\tilde{b}(t) -\left(\kappa +\kappa_{L}\right)
\tilde{a}(t)+\sqrt{2\kappa}\tilde{a}^{in}(t)
+\sqrt{2}\bar{\kappa}_{L}^*\tilde{a}_{+}^{in}(t)+F_a, \\
\label{QLE:b2nd}
\dot{\tilde{b}}(t)&=&i\left(\nu^{\prime}-\nu-\delta_2^b\right)
\tilde{b}(t) -\bar{\chi}_2^* \tilde{a}(t)
-\left(\kappa_b+\kappa_{+2}^{b}-\kappa_{-2}^b\right)
\tilde{b}(t)+\sqrt{2\kappa_b}\tilde{b}^{in}(t)
+\sqrt{2}\bar{\kappa}_{+2}^b\tilde{a}_{+}^{in}(t)-\sqrt{2}\bar{\kappa}_{-2}^b\tilde{a}_{-}^{in}(t)^{\dagger}+F_b.
\end{eqnarray} \end{widetext} Let us now define the coefficients
appearing in these equations. The quantities $\kappa_L$ and
$\delta'$ are given by Eq.~(\ref{los1}) and (\ref{eqfornu1}),
respectively. In fact, together with $\bar{\kappa}_L$, they do not
depend upon the properties of the driving laser; moreover the two
noise operators $\tilde{a}_{-}^{in}(t)$ and
$\tilde{a}_{+}^{in}(t)$ are given by Eqs.~(\ref{nos1})
and~(\ref{nos2}). The nonlinear terms $F_a$ and $F_b$ are the same
as in Eqs.~(\ref{QLE:a1})-(\ref{QLE:b}), and are negligible as we
take $\Omega_2\cos\theta_L\gg g_c\cos\theta_c$. The coupling
constants associated with the Raman scattering processes are
\begin{eqnarray} &&\chi_2=\eta \Omega_2
g_c^*\cos\phi_c\left(\frac{\cos\theta_L}{\Delta_2+\nu^{\prime}+{\rm
i}\gamma/2}
+\frac{{\rm i}\tan\phi_c\cos\theta_c}{\Delta_2+{\rm i}\gamma/2}\right)~,\nonumber\\
&&\label{Chi:2} \\
&&\bar{\chi}_2=\eta \Omega_2
g_c^*\cos\phi_c\left(\frac{\cos\theta_L}{\Delta_2+\nu^{\prime}-{\rm
i}\gamma/2}
+\frac{{\rm i}\tan\phi_c\cos\theta_c}{\Delta_2+{\rm i}\gamma/2}\right)~.\nonumber\\
&&\label{Chi:2:t} \end{eqnarray} As we consider the limit $\gamma
\ll |\Delta_2|$ we shall take $\bar{\chi}_2=\chi_2$ from now on.
The incoherent emission or absorption of a vibrational quantum
scales with the rates \begin{eqnarray}
\kappa_{\pm2}^{b}=\frac{\gamma}{2}\frac{\eta^2\cos^2\theta_L\Omega_2^2}{\gamma^2/4+(\Delta_2\pm\nu^{\prime})^2},
\end{eqnarray} and in the limit $|\Delta_2|\gg \gamma,\nu$ the
rates scaling the input noise read \begin{eqnarray*}
&&\bar{\kappa}_{-2}^b=-i\eta\sqrt{\frac{\gamma}{2}}\frac{\Omega_2\cos\theta_L}{\gamma/2+i(\Delta_2-\nu'')}\\
&&\bar{\kappa}_{+2}^b=i\eta\sqrt{\frac{\gamma}{2}}\frac{\Omega_2^*\cos\theta_L}{\gamma/2-i(\Delta_2+\nu'')}.
\end{eqnarray*}
Finally, the frequency shift of the vibrational motion reads
\begin{eqnarray} &&\delta_{2}^{b}= \frac{2\Delta_2 \eta^2
|\Omega_2|^2 \cos^2\theta_L\left(\gamma^2/4+\Delta_2^2-\nu^{\prime
2}\right)}{\left(\gamma^2/4+\Delta_2^2-\nu^{\prime 2}\right)^2
+\nu^{\prime 2}\gamma^2}\label{nu2b} \\
&&~-\eta^2 |\Omega_2|^2
\cos^2\theta_L\frac{2\Delta_2}{\Delta_2^2+\gamma^2/4}\nonumber
\end{eqnarray}
For $|\Delta_2|\gg\gamma$, $\eta|\Omega_2/\Delta_2|\ll\nu$, we
find with good approximation \begin{eqnarray}\label{renfreq2}
  \delta_{2}^{b}&\approx& \frac{2\Delta_2 \eta^2 |\Omega_2|^2
  \cos^2\theta_L\left(\gamma^2/4+\Delta_2^2-\nu^2\right)}{\left(\gamma^2/4+\Delta_2^2-\nu^2\right)^2+\nu^2\gamma^2}\nonumber\\
& &~-\eta^2 |\Omega_2|^2
\cos^2\theta_L\frac{2\Delta_2}{\Delta_2^2+\gamma^2/4},
\end{eqnarray}
determining, together with Eq.~(\ref{eqfornu1}), the resonance
condition for the central frequency of the laser pulse,
\begin{equation} \delta_2 = \omega_{L2}-\omega_c=\delta'
-\nu-\delta_{2}^{b} \label{res12} . \end{equation} For
$|\Delta_2|\gg\nu$, choosing $\Omega_2=\Omega_1$, then
$\delta_2^b\approx\delta_1^b$, $\nu^{\prime\prime}=\nu^{\prime}$
and the processes leading to absorption of a phonon and emission
of a cavity photon are resonantly enhanced by choosing the
frequency of the second laser pulse at
$\omega_{L2}=\omega_c-\nu^{\prime}$. We consider this regime, as
it simplifies substantially the calculations. When it is not
fulfilled, one must consider an accumulated phase, which gives
simply a total phase shift and hence modifies the quadratures
exhibiting entanglement.

\section{Calculation of the elements of the correlation matrix}
\label{App:B}

We derive here the elements of the intracavity correlation matrix
$V$ of Eq.~(\ref{correv}), using that
the cavity modes and the vibrational motion are initially in the
vacuum state, and the fact that input noise is uncorrelated with the
cavity mode operators at former times. The elements read
\begin{eqnarray}
V_{12} &=& 0\\
V_{11} &=& V_{22}= \left\langle \tilde{a}^{\dag}(T_1)\tilde{a}(T_1)\right\rangle +1/2\nonumber\\
V_{33} &=& \left\langle \tilde{a}^{\dag}(T_2)\tilde{a}(T_2)\right\rangle +{\rm Re}\left\{\langle \tilde{a}^{\dag}(T_2)^2\rangle \right\}+1/2 \nonumber\\
V_{44} &=& \left\langle \tilde{a}^{\dag}(T_2)\tilde{a}(T_2)\right\rangle -{\rm Re}\left\{\langle \tilde{a}^{\dag}(T_2)^2\rangle \right\}+1/2 \nonumber\\
V_{34} &=& -{\rm Im}\left\{\langle \tilde{a}^{\dag}(T_2)\tilde{a}^{\dag}(T_2)\rangle \right\}\nonumber\\
V_{13} &=& \left[\left\langle \tilde{a}(T_1)\tilde{a}(T_2)\right\rangle+\left\langle\tilde{a}(T_1)\tilde{a}^{\dag}(T_2)\right\rangle\right.\nonumber\\
&&\left.+\left\langle\tilde{a}^{\dag}(T_1)\tilde{a}(T_2)\right\rangle+\left\langle\tilde{a}^{\dag}(T_1)\tilde{a}^{\dag}(T_2)\right\rangle+c.c.\right] /4\nonumber\\
V_{24} &=& -\left[\left\langle \tilde{a}(T_1)\tilde{a}(T_2)\right\rangle-\left\langle\tilde{a}(T_1)\tilde{a}^{\dag}(T_2)\right\rangle\right.\nonumber\\
&&\left.-\left\langle\tilde{a}^{\dag}(T_1)\tilde{a}(T_2)\right\rangle+\left\langle\tilde{a}(T_1)^{\dag}\tilde{a}^{\dag}(T_2)\right\rangle+c.c. \right]/4\nonumber\\
V_{14} &=& -i\left[\left\langle \tilde{a}(T_1)\tilde{a}(T_2)\right\rangle-\left\langle \tilde{a}(T_1)\tilde{a}^{\dag}(T_2)\right\rangle\right.\nonumber\\
&&\left.+\left\langle \tilde{a}^{\dag}(T_1)\tilde{a}(T_2)\right\rangle-\left\langle
\tilde{a}^{\dag}(T_1)\tilde{a}^{\dag}(T_2)\right\rangle-c.c.\right] /4\label{correl} \nonumber \\
V_{23} &=& -i\left[\left\langle \tilde{a}(T_1)\tilde{a}(T_2)\right\rangle+\left\langle \tilde{a}(T_1)\tilde{a}^{\dag}(T_2)\right\rangle\right.\nonumber\\
&&\left.-\left\langle \tilde{a}^{\dag}(T_1)\tilde{a}(T_2)\right\rangle-\left\langle
\tilde{a}^{\dag}(T_1)\tilde{a}^{\dag}(T_2)\right\rangle-c.c.\right] /4\label{correlb} \nonumber
\end{eqnarray}
Using Eqs.~(\ref{asol1}), (\ref{qlesol1b}) and (\ref{asol2}) we obtain
\begin{widetext} \begin{eqnarray} \left\langle
\tilde{a}^{\dag}(T_1)\tilde{a}(T_1)\right\rangle
&=&\left|\chi_1\right|^2\left(|f_1(T_1)|^2+(\kappa_h+2|\bar{\kappa}_{+1}^{b}|^2)\int^{T_1}_{0}{\rm d}s\:|f_1(s)|^2\right)\\
\left\langle \tilde{a}^{\dag}(T_2)\tilde{a}(T_2)\right\rangle
&=&|g_{-2}(T_2-T)|^2e^{-2(\kappa+\kappa_L)(T-T_1)}\left\langle
\tilde{a}^{\dag}(T_1)\tilde{a}(T_1)\right\rangle
+\left|\chi_2\right|^2|f_2(T_2-T)|^2e^{-2\kappa_b(T-T_1)}\left\langle
\tilde{b}^{\dag}(T_1)\tilde{b}(T_1)\right\rangle\nonumber\\
&+&|\chi_2|^2\left(\kappa_h+2|\bar{\kappa}_{-2}^b|^2\right)\int_0^{T_2-T}|f_2(s)|^2
+\frac{\kappa_h}{2\kappa_b}\left|\chi_2\right|^2|f_2(T_2-T)|^2\left(1-e^{-2\kappa_b(T-T_1)}\right)\\
\left\langle \tilde{a}(T_1)\tilde{a}(T_2)\right\rangle&=&
\chi_2f_2(T_2-T){\rm e}^{(i\delta_1^b-\kappa_b)(T-T_1)}
\left\langle \tilde{a}(T_1)\tilde{b}(T_1)\right\rangle
\\
\left\langle
\tilde{a}^{\dag}(T_1)\tilde{a}^{\dag}(T_2)\right\rangle&=&
=\left\langle \tilde{a}(T_1)\tilde{a}(T_2)\right\rangle^*\\
\left\langle
\tilde{a}^{\dag}(T_2)\tilde{a}^{\dag}(T_2)\right\rangle&=&
2\chi_2^*f_2(T_2-T)g_{-2}(T_2-T){\rm
e}^{-(i\delta_1^b+\kappa_b+\kappa_L+\kappa)(T-T_1)}\left\langle
\tilde{a}^{\dag}(T_1)\tilde{b}^{\dag}(T_1)\right\rangle \\
\left\langle
\tilde{a}(T_1)\tilde{a}^{\dag}(T_2)\right\rangle&=&g_{-2}(T_2-T)e^{-(\kappa_L+\kappa)(T-T_1)}
\left\langle \tilde{a}(T_1)\tilde{a}^{\dag}(T_1)\right\rangle\\
\left\langle
\tilde{a}^{\dag}(T_1)\tilde{a}(T_2)\right\rangle&=&\left\langle
\tilde{a}(T_1)\tilde{a}^{\dag}(T_2)\right\rangle^* \end{eqnarray}
\end{widetext} where we have used $\kappa_h=\kappa_b(2\bar{N}+1)$,
and that $\kappa_h\gg\kappa_b$. In deriving these relation we also
used that
$\langle\tilde{a}^{\dagger}(T_1)\tilde{b}(T_1)\rangle=\langle\tilde{a}(T_1)\tilde{b}^{\dag}(T_1)\rangle=0$.
In order to fully determine the above relations as a function of
the initial conditions, we need \begin{widetext} \begin{eqnarray}
&&\left\langle
\tilde{b}^{\dag}(T_1)\tilde{b}(T_1)\right\rangle=|\chi_1|^2|f_1(T_1)|^2+\kappa_h\int_0^{T_1}{\rm
d}s~|g_{+1}(s)|^2+2\kappa|\chi_1|^2\int_0^{T_1}{\rm
d}s~|f_{1}(s)|^2
\\
&&+2\int_0^{T_1}{\rm d}s~\left|\bar{\kappa}_{-1}^{b}g_{+1}(s)-\chi_1\bar{\kappa}_Lf_{1}(s)\right|^2\;\nonumber\\
&&\left\langle
\tilde{a}(T_1)\tilde{b}(T_1)\right\rangle=g_{-1}(T_1)\chi_1f_1(T_1)+\chi_1(\kappa_h-2\kappa_b)\int_0^{T_1}{\rm
d}s~f_1(s)g_{+1}(s)
\\
&&+2\int_0^{T_1}{\rm d}s~\left(\bar{\kappa}_{-1}^{b*}\chi_1
f_1(s)-\bar{\kappa}_L^*g_{-1}(s)\right)
\left(\bar{\kappa}_{-1}^{b}g_{+1}(s)-\bar{\kappa}_L\chi_1f_{1}(s)\right)
+2\kappa\chi_1\int_0^{T_1}g_{-1}(s)f_1(s)\nonumber
\end{eqnarray}

\end{widetext}

In deriving these expressions we used that the HLE conserve the commutation relations.
This is not fulfilled for long times, when the perturbative expansion loses validity.

\end{appendix}


\begin{thebibliography}{99}
\bibitem{Footnote} Note that Eqs.~(\ref{b_c}) and~(\ref{b_L}) are
at first order in the Lamb-Dicke parameter, nevertheless we have
included also the term $\eta^2(2b^{\dagger}b+1)$, which belongs to
the second-order expansion. This term gives rise to a.c.-Stark
shifts induced by the field, which should be systematically taken
into account when one considers transition rates between
vibrational states (which are at second order in $\eta$). See for
instance \protect\cite{Eschner03,Vitali06}.
\end{thebibliography}
\end{document}